\documentclass[floatfix, twocolumn,amsmath,amssymb, aps,prb,superscriptaddress]{revtex4-2}
\usepackage{mathrsfs}
\usepackage{amsmath}
\usepackage{ragged2e}
\usepackage{graphicx}
\usepackage{caption}
\usepackage{subcaption}
\usepackage{hyperref}
\usepackage{stfloats}
\usepackage{ulem}
\hypersetup{ colorlinks, citecolor=blue, linkcolor=blue, urlcolor=blue}

\begin{document}
	\title{Vanishing of the quantum spin Hall phase in a semi-Dirac Kane Mele model}
	
	\author{Sayan Mondal}
	\affiliation{Department of Physics, Indian Institute of Technology Guwahati, 
		Guwahati 781039, Assam, India}
	\author{Saurabh Basu}
	\affiliation{Department of Physics, Indian Institute of Technology Guwahati, 
		Guwahati 781039, Assam, India}

	\begin{abstract}
		We study the vanishing of the topological properties of a quantum spin Hall insulator induced by a deformation of the band structure that interpolates between the Dirac and the semi-Dirac limits of a tight-binding model on a honeycomb lattice. The above scenario is mimicked in a simple model, where there exists a differential hopping along one of the three neighbours (say, $t_1$) compared to the other two (say, $t$). For $t_1 = t$, the properties of the quantum spin Hall phase is described by the familiar Kane Mele model, while $t<t_1<2t$ denotes a situation in which the spin resolved bands are continuously deformed. $t_1 = 2t$ represents a special case which is called as the semi-Dirac limit. Here, the spectral gaps between the conduction and the valence bands vanish. A closer inspection of the properties of such a deformed system yields insights on a topological phase transition occurring at the semi-Dirac limit, which continues to behave as a band insulator for $t_1>2t$. We demonstrate the evolution of the topological phase in presence of the Rashba and intrinsic spin-orbit couplings via computing the electronic band structure, edge modes in a nanoribbon  and the $\mathbb{Z}_2$ invariant. The latter aids in arriving at the phase diagram which conclusively shows vanishing of the topological phase in the semi-Dirac limit. Further we demonstrate in gradual narrowing down of the plateau in the spin Hall conductivity, which along with a phase diagram provide robust support on the vanishing of the $\mathbb{Z}_2$ invariant and hence the quantum spin Hall phase.
		
	\end{abstract}
	\maketitle
	
	\section{Introduction}
	Engineering topological phases in materials has been a topic of intense scrutiny in studies of condensed matter systems. The realization of the topological phases of matter became evident after the discovery of quantum Hall effect (QHE) \cite{thouless1982}, which tells us that in presence of a strong perpendicular magnetic field, the Hall conductivity is quantized in unit of $e^2/h$. This quantization occurs because of the presence of magnetic Bloch bands owing to the presence of the magnetic field. The quantized Hall conductivity is associated with a non-zero integer, known as Thouless-Kohmoto-Nightingale-Nijs (TKNN) invariant.
	
	Just when an external magnetic field is thought to be indispensable for the realization of a quantum Hall effect, it was Haldane \cite{Haldane1988} who proposed that QHE can be obtained even in absence of the external magnetic field. He introduced a complex next nearest neighbour (NNN) hopping in a honeycomb lattice, a prototype for graphene. Such a hopping breaks the time reversal symmetry (TRS), which is understood to be more fundamental for observing QHE than an external field. It was shown that the bands posses a non-zero Chern number which yields the quantized Hall conductivity similar to the QHE when the Fermi energy lies in the band gap. In the bandstructure of the Haldane model, the closing and opening of the energy gap occurs at the Dirac points ($\mathbf{K}$ and $\mathbf{K}^\prime$) depending on the value of the Semenoff \cite{semenhoff} mass. The variation of the Semenoff mass as a function of the Haldane flux, $\phi$ demonstrates a phase diagram \cite{Haldane1988,vanderbilt2006} that involves opening and closing of the spectral gaps alternately at the $\mathbf{K}$ and the $\mathbf{K}^\prime$ points.
	
	Since the observation of QHE demands violation of TRS, a natural question is what happens when TRS is intact. The Chern number must vanish in such systems. Considering this scenario, Kane and Mele \cite{kanemele_2005} have proposed a spin-full model, where two copies of the Haldane model have been considered. In these copies, they have added a positive Haldane flux for spin-$\uparrow$ electrons and negative one for the spin-$\downarrow$ electrons. Kane and Mele found that the Chern numbers for different spin bands are same, but are of opposite signs and hence the total Chern number vanishes, which is consistent with the requirement of TRS. However, the difference of those Chern numbers is non-zero which means the system can still be defined by a non-zero topological invariant, albeit the Hall effect in the charge sector vanishes. This type of insulators fall into the class of $\mathbb{Z}_2$ topological insulator which has non-zero $\mathbb{Z}_2$ index. The non-zero $\mathbb{Z}_2$ invariant is associated with a finite \textit{spin} Hall conductivity \cite{kanemele_2005_2}. Further, Kane and Mele have also included a $S_z$ symmetry breaking Rashba \cite{rashba1960} spin-orbit coupling (SOC) term, which is a reasonable assumption for a system without surface inversion symmetry (indeed a situation for graphene). Further they have shown that  there emerges a  phase diagram  that encodes topological phase transition where the $\mathbb{Z}_2$ invariant shows a discontinuous jump from a value 1 to 0. The system shows quantized spin Hall conductivity if the Fermi energy lies in the bulk gap. Experimentally, a number of materials have been found that show spin Hall conductivity, such as, CdTe-HgTe \cite{bernevig2006},  Pt \cite{kimura2007}, Cl-doped ZnSe \cite{stern2006}, GaAs \cite{zhao2006}, FePt/Au devices \cite{seki2008}, lateral spin valve structure, namely Pt/Au \cite{isasa2015} etc. A non-zero spin Hall conductivity yields a feasible proposal of manipulating the spin degree of freedom for dissipationless transport, and thus provide impetus to the field of spintronics.
	
	In this paper, we focus on the Kane Mele model that includes the Rashba spin-orbit term, in addition to the complex NNN hopping with different (same magnitude, but opposite signs) Haldane fluxes for the two spins. The latter is referred to as the intrinsic SOC.
	In this paper, we simulate the topological properties of the Kane Mele model and superpose the effects of deformation of the band structure by inducing an anisotropy in the hopping energies on a honeycomb lattice. Such an anisotropy shifts the band minima points, which eventually merge onto a single point in the semi-Dirac limit ($t_1 = 2t$), intervening the two Dirac points in the BZ. Such a shift and subsequent merger of the band minima or the Dirac points \cite{montambaux2009,zieglar2017,hasegawa2006}, and their roles in inducing topological phase transitions have been discussed in the literature for spinless systems \cite{mondal2021}. However, to the best of our knowledge, such shift and merger of the spin bands have not been studied in spin-full systems. Particularly, the band deformation effects are interesting to study in the context of spin transport and evolution of the topological phases in systems where the TRS is intact. Motivated by the above scenario, here we have explored the evolution of the topological properties of the band deformed quantum spin Hall insulators. 
	
	Let us include a short note on the band deformation and their implications on the topological properties. Among a large number of possibilities of deforming the band structure, of which the experimentally realized variants are germanene and stanene \cite{ni2017}, different optimized structures \cite{schroter2017,wu2016}, such as, $\alpha$-graphyne, $\beta$-graphyne \cite{baughman1987} etc, we choose one of the three hopping frequencies (say, $t_1$) to be different than the other two (say, $t$). In a particular situation, where $t_1$ equals $2t$, the low energy Hamiltonian demonstrates a quadratic dispersion along one direction in the BZ, and a Dirac-like along the other. The corresponding scenario is known as the semi-Dirac case. In a general sense, the strength $t_1$ may be tuned from $t$ to any value, say $t_1\geq 2t$. In the range $t\leq t_1 < 2t$, we get the QSH phase with a non-zero $\mathbb{Z}_2$ number. An important feature of this regime is that the band extrema in the electronic spectrum move from the so called Dirac points ($\mathbf{K}$ and $\mathbf{K}^\prime$) towards each other and merge into a single band touching point, called as the $\mathbf{M}$ point and the system respects time reversal invariance all the while. At the $\mathbf{M}$ point, in the low energy limit, the band structure displays anisotropic linear dispersion, in the sense, the velocities are different along the different directions. We shall see that the $\mathbb{Z}_2$ invariant vanishes as soon as $t_1$ becomes equal to $2t$, where a trivial spin Hall insulating phase emerges, thereby indicating a transition from a topological phase to a trivial one through a gap closing point. Thus it remains to be seen how the $\mathbb{Z}_2$ number phase diagram vanishes and the spin Hall conductance responds to the tuning of the band structure across the gap closing critical point. To address such questions, we perform a detailed study  of the evolution of band structure, edge modes in a nanoribbon, the phase diagram and the spin Hall conductivity to study the evolution of topological properties. 
	
	The paper is organized as follows, in section \ref{sec:modelhamiltonian} we show our Kane Mele model Hamiltonian in presence of a hopping anisotropy inside a unit cell on a honeycomb lattice. In section \ref{sec:edgestates}, we have depicted the evolution of the edge states in a nanoribbon and specifically depicted the edge modes and the currents carried by them in relevance to the ongoing discussion. Then, in section \ref{sec:phasediagram} we investigate the evolution of the phase diagram which characterizes the topological phase and also its disappearance in a quantitative manner. Finally we show the spin Hall conductivity  in section \ref{sec:spinhallcond} and conclude with a brief summary of our results in section \ref{sec:conclusion}.
	
	\section{Model Hamiltonian}\label{sec:modelhamiltonian}
\begin{figure}[h]
		\centering
		\includegraphics[width=0.4\textwidth]{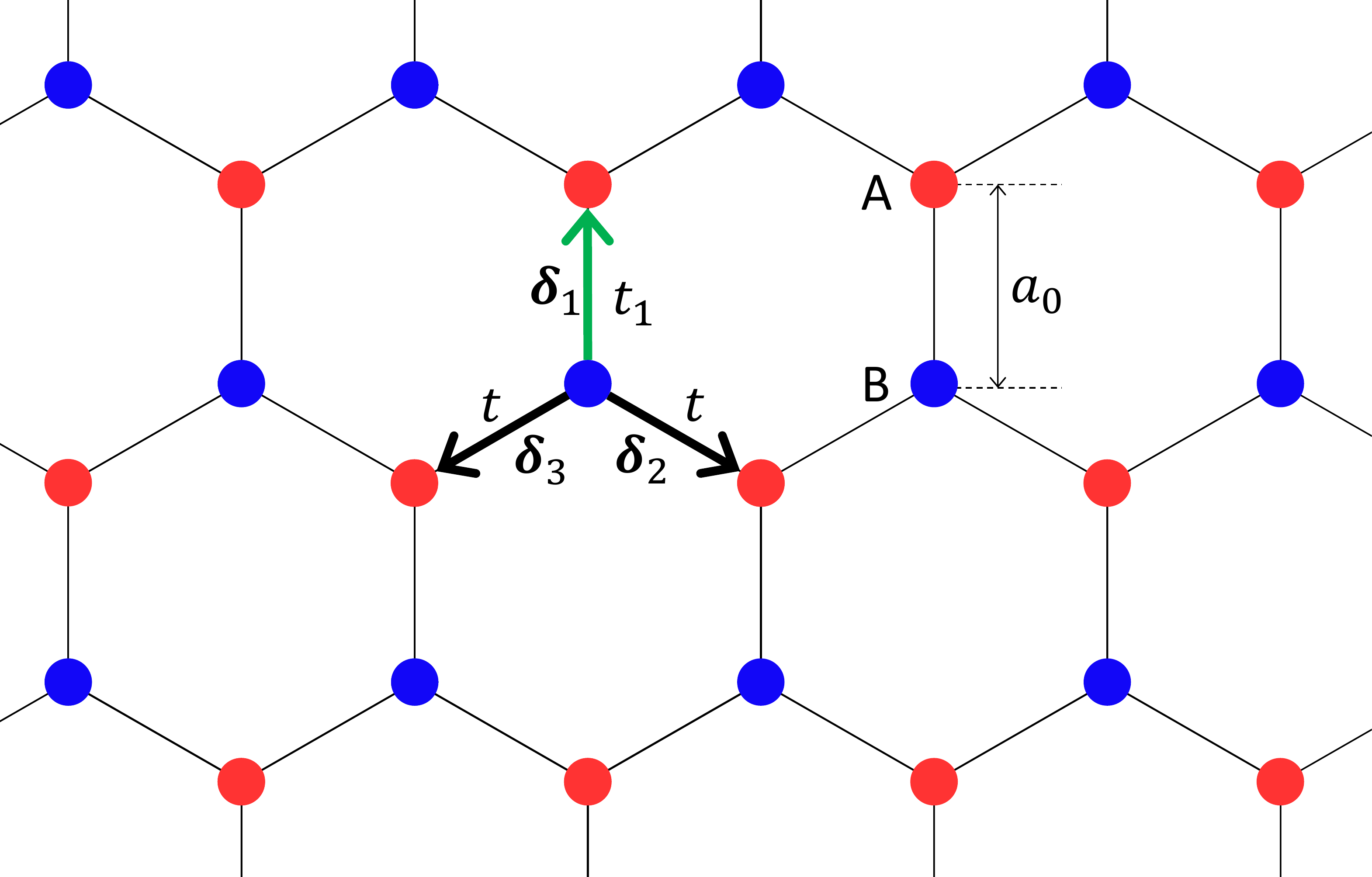}
		\caption{\raggedright A honeycomb lattice is shown where the red and the blue circles represent the sublattices A and B respectively. In the $\boldsymbol{\delta}_2$ and $\boldsymbol{\delta}_3$ directions the NN hopping strengths are same ($t$) while in the $\boldsymbol{\delta}_1$ direction it is $t_1$.}\label{honeycomb}
	\end{figure}

	Consider the Kane Mele Hamiltonian on a honeycomb lattice with different nearest neighbour (NN) hopping strengths which we show in Fig. \ref{honeycomb}. The hopping strengths along the $\boldsymbol{\delta}_2$ and $\boldsymbol{\delta}_3$ directions are $t$, while in the third direction, that is along $\boldsymbol{\delta}_1$, the strength is $t_1$. The NN vectors are given by $\boldsymbol{\delta}_1 = a(0, 1)$, $\boldsymbol{\delta}_2= a(\sqrt{3}/2, -1/2)$ and $\boldsymbol{\delta}_3= a(-\sqrt{3}/2, -1/2)$.  We assume $t_1$ is a tunable parameter and is defined in the range, $t\leq t_1\leq 2t$. The value $t_1 = 2t$ denotes a critical point which we termed as the semi-Dirac case. We have also considered values beyond the semi-Dirac limit, namely $t_1>2t$. The Hamiltonian can be written as follows,
	
	\begin{align}\label{ham1}
		H =& \sum_{\langle ij \rangle} t_{ij} c_i^\dagger c_j + i \lambda_{\mathrm{SO}}\sum_{\langle\langle ij \rangle\rangle} \nu_{ij} c_i^\dagger s^z c_j \nonumber\\ 
		&+ i \lambda_\mathrm{R} \sum_{\langle ij \rangle} c_i^\dagger (\mathbf{s} \times \hat{\mathbf{d}}_{ij})_z c_j + \lambda_\mathrm{v} \sum_i \xi_i c_i^\dagger c_j
	\end{align}
	Here $t_{ij} = t_1$ or $t$ when $j$ denotes the neighbours $\boldsymbol{\delta}_{1}$ or $\boldsymbol{\delta}_{2,3}$ respectively. The first term denotes the direction dependent nearest neighbour hopping. The second term is the complex NNN hopping which encodes a sign difference between the spin-$\uparrow$ and spin-$\downarrow$ electrons. It is usually denoted as the intrinsic spin-orbit coupling with a coupling strength $\lambda_{\mathrm{SO}}$. Here, $\nu_{ij} = \mathrm{sgn}(\hat{\mathbf{d}}_1 \times \hat{\mathbf{d}}_2)_z = \pm 1$, where $\hat{\mathbf{d}}_1$ and $\hat{\mathbf{d}}_2$ are the unit vectors along the two bonds that electron traverses while going from site $j$ to NNN site $i$. $\mathbf{s}$ denotes the Pauli matrices describing the electron spin. The third term is the nearest neighbour Rashba spin-orbit interaction with strength $\lambda_{\mathrm{R}}$ where $\hat{\mathbf{d}}_{ij}$ is a unit vector of nearest neighbour sites pointing from site $i$ to site $j$. This term breaks the $z\rightarrow -z$ mirror symmetry, and as a result, the spins in the $z$ direction are no longer conserved. The fourth term is the staggered sublattice potential that induces an energy difference between A and B sublattices and $\xi_i = \pm 1$. It is often referred to as the Semenoff mass. 
	
	Before we solve the above Hamiltonian, it is instructive to mention that the low energy version of the above Hamiltonian has a form $\tau_z \sigma_z s_z$, where $\tau_z$ denotes the two valleys and assumes values $\pm 1$ for the them, $\sigma_z$ is the sublattice (A and B) degree of freedom and $s_z$ denotes spin. Under time reversal operation, the valleys and the spins change their signs, while nothing happens to the sublattices. This implies, under time reversal, $\tau_z \rightarrow -\tau_z$, $\sigma_z \rightarrow \sigma_z$, $s_z\rightarrow -s_z$ and hence the Hamiltonian possesses TRS. 
	
	Performing a Fourier transform on the Hamiltonian in Eq. \ref{ham1} can be written as,
	
	\begin{align}\label{ham_kspace}
		H&(\mathbf{k}) = \nonumber\\
		&\begin{pmatrix}
			\gamma(\mathbf{k}) + \lambda_\mathrm{v} & \zeta(\mathbf{k}) & 0 & \rho(\mathbf{k}) \\ 
			\zeta^*(\mathbf{k}) & -\gamma(\mathbf{k}) - \lambda_\mathrm{v} & -\rho(-\mathbf{k}) & 0 \\
			0 & -\rho^*(-\mathbf{k}) & -\gamma(\mathbf{k}) + \lambda_\mathrm{v} & \zeta(\mathbf{k}) \\
			\rho^*(\mathbf{k}) & 0 & \zeta^*(\mathbf{k}) & \gamma(\mathbf{k}) - \lambda_\mathrm{v}	
		\end{pmatrix}
	\end{align}
	where,
	\begin{subequations}
		\begin{equation}
			\zeta(\mathbf{k}) = t_1 e^{-i k_y a}  + 2 t e^{\frac{i k_y a}{2}} \cos\frac{\sqrt{3}k_x a}{2}
		\end{equation}
		\begin{equation}
			\gamma(\mathbf{k}) = 2\lambda_{\mathrm{SO}}\left[2 \sin\frac{\sqrt{3} k_x a}{2} \cos\frac{3 k_y a}{2} - \sin\sqrt{3} k_x a\right]
		\end{equation}
		$\mathrm{and,}$
		\begin{equation}
			\rho(\mathbf{k}) = i \lambda_\mathrm{R}\left[ e^{-i k_y a}  + e^{\frac{i k_y a}{2}} 2\cos\left( \frac{\sqrt{3}k_x a}{2} + \frac{\pi}{3}  \right) \right]
		\end{equation}
		
	\end{subequations}
	\begin{figure}[h]
		\begin{center}
			\begin{subfigure}[b]{0.235\textwidth}
				\includegraphics[width=\textwidth]{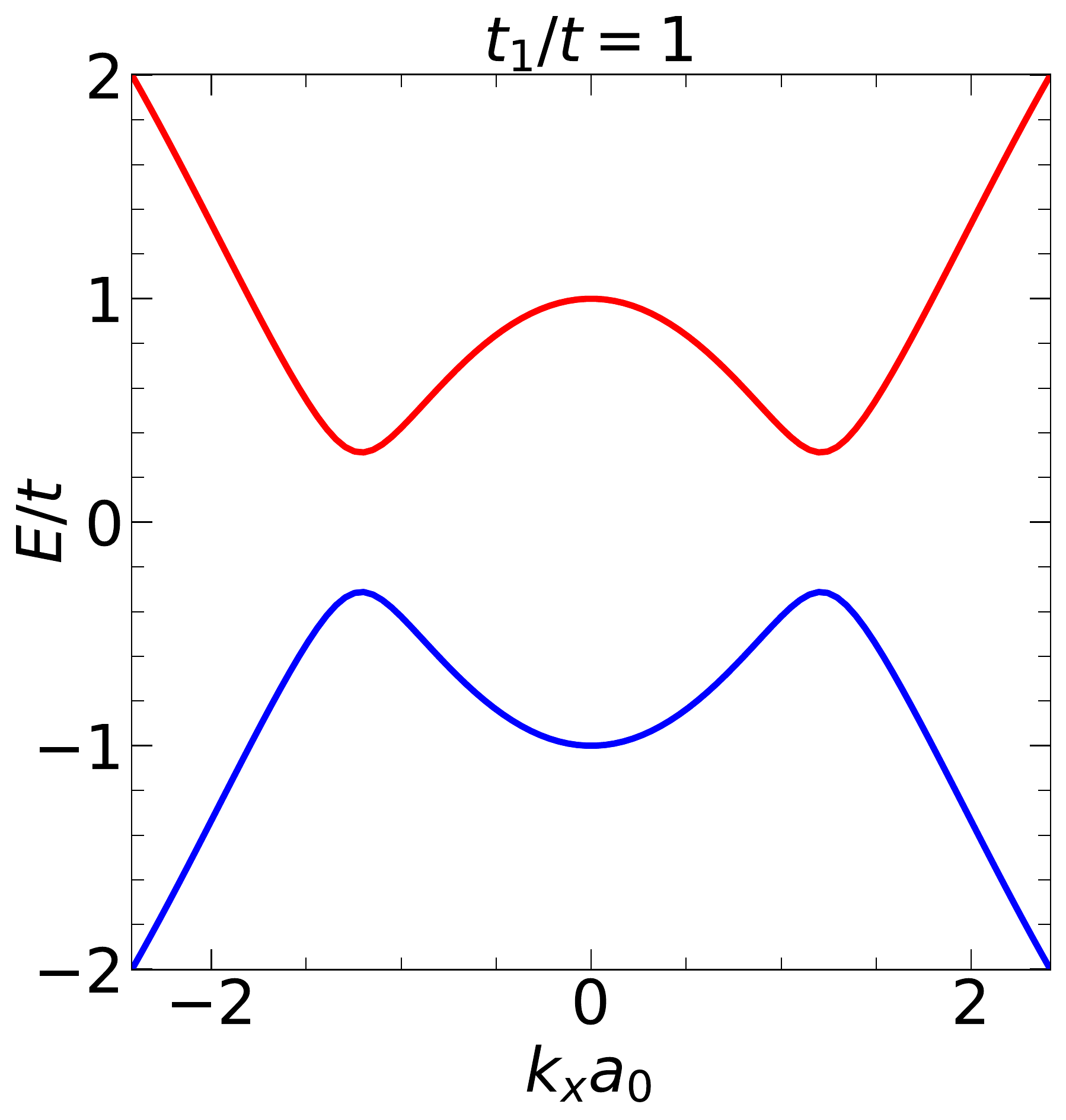}
				\subcaption{}\label{band5}
			\end{subfigure}
			\begin{subfigure}[b]{0.235\textwidth}
				\includegraphics[width=\textwidth]{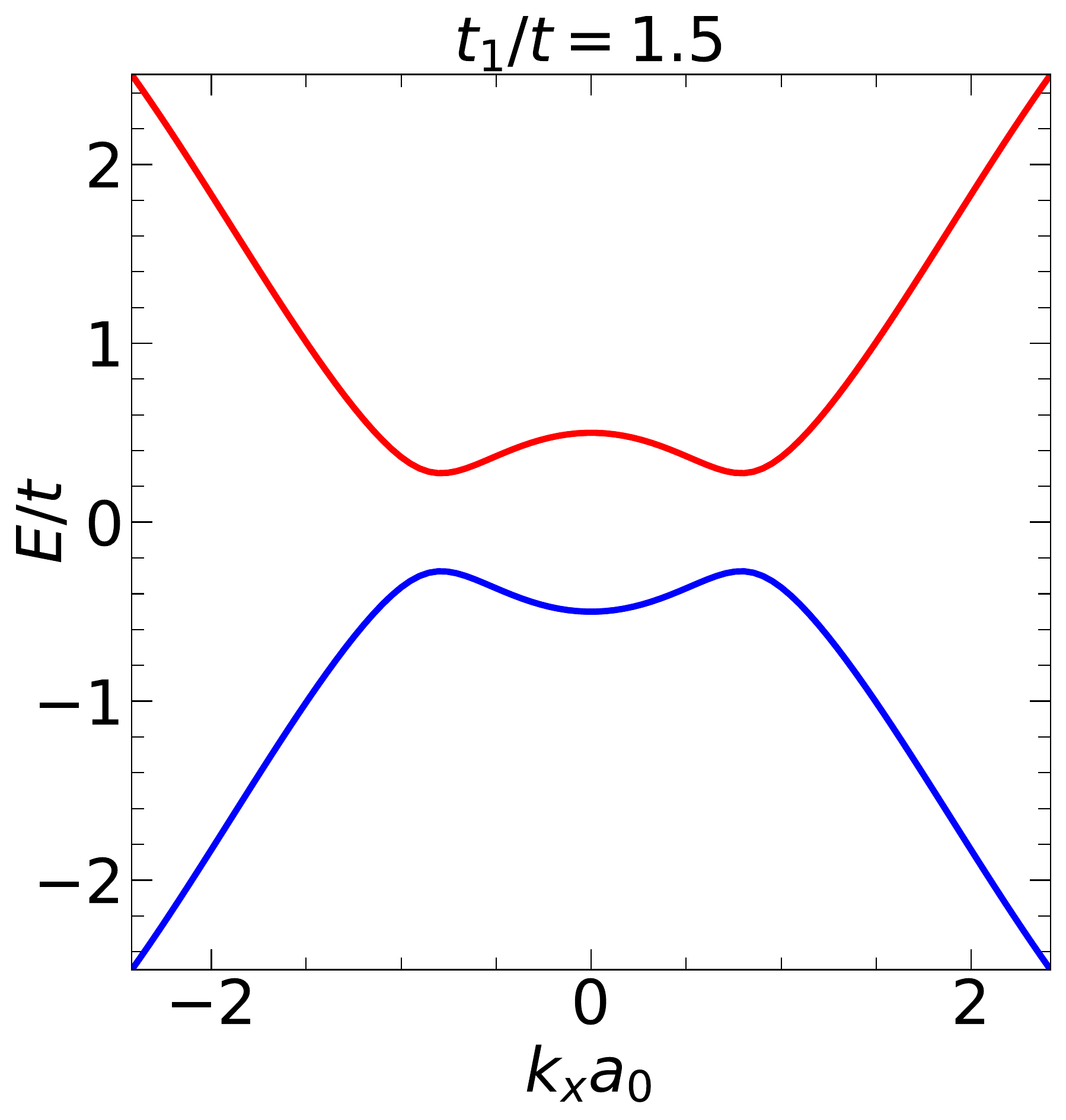}
				\subcaption{}\label{band6}
			\end{subfigure}
			\begin{subfigure}[b]{0.235\textwidth}
				\includegraphics[width=\textwidth]{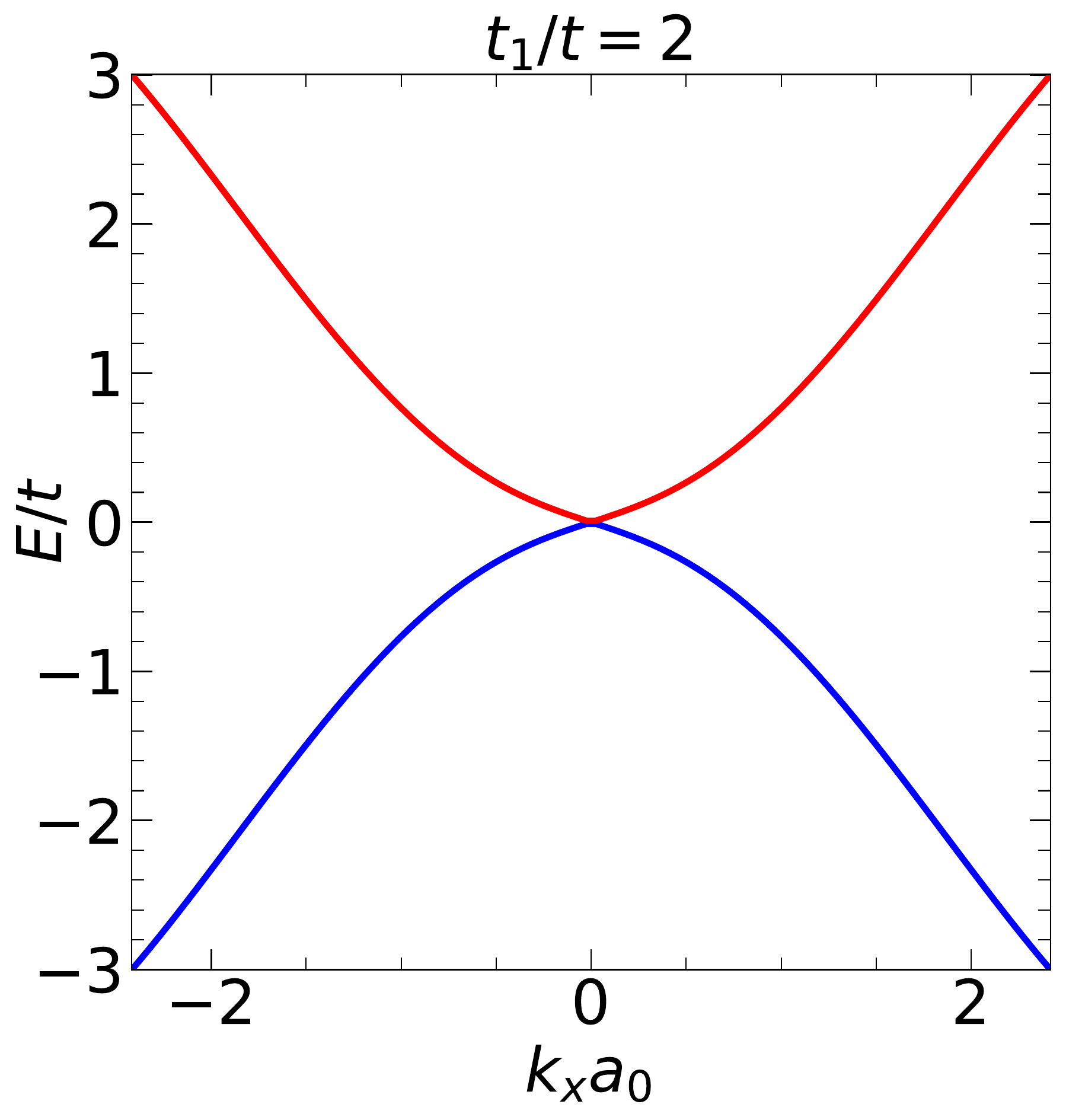}
				\subcaption{}\label{band7}
			\end{subfigure}
			\begin{subfigure}[b]{0.235\textwidth}
				\includegraphics[width=\textwidth]{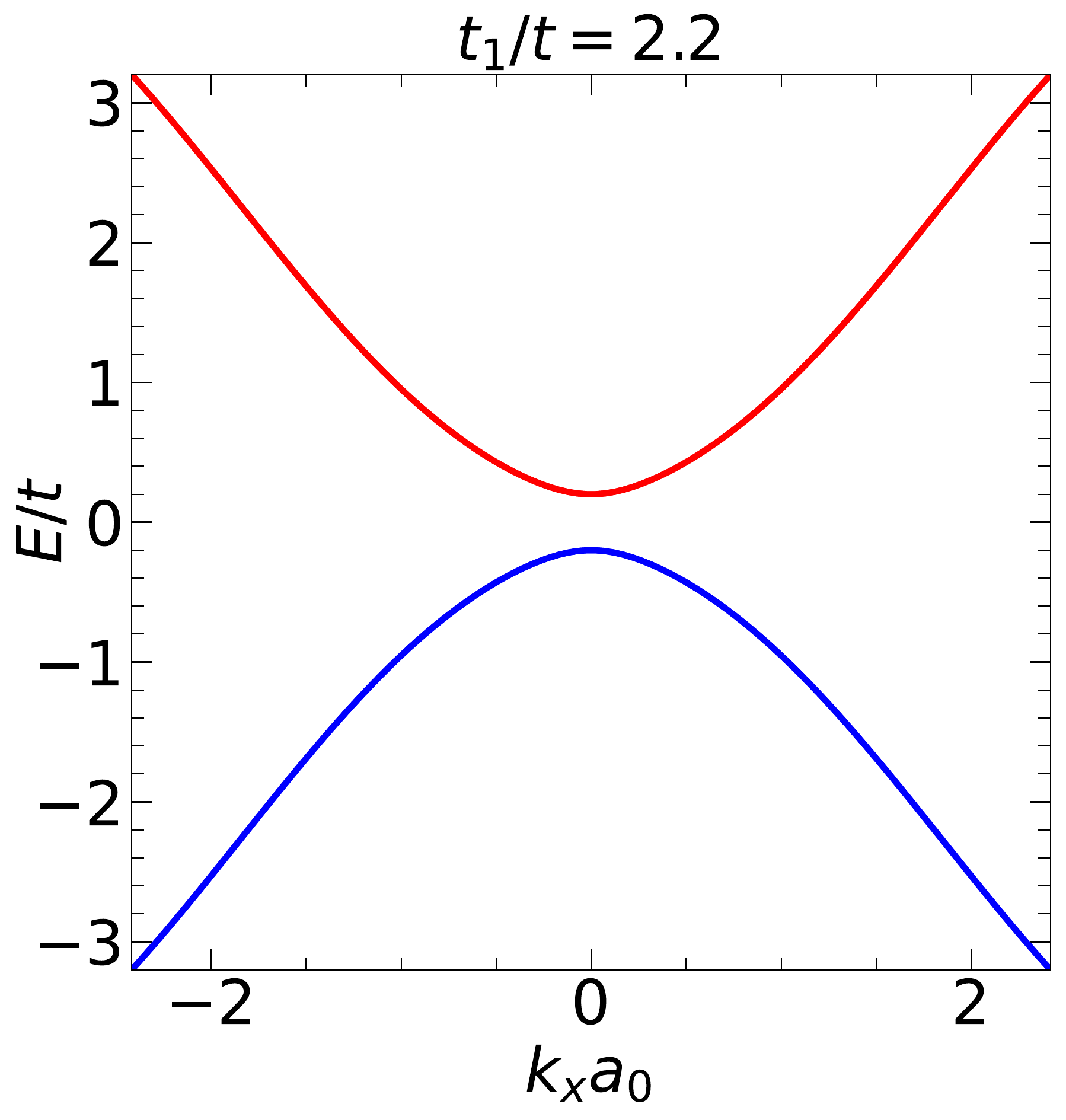}
				\subcaption{}\label{band8}
			\end{subfigure}
			\caption{\raggedright The energy band structure corresponding to the Hamiltonian in Eq. (\ref{ham_kspace}) along the $k_x$ axis ($k_y a_0$ is fixed at $2\pi/3$) are shown for (a) $t_1/t = 1$, (b) $t_1/t = 1.5$, (c) $t_1/t = 2$ and (d) $t_1/t = 2.2$ in absence of Rashba SOC ($\lambda_{\mathrm{R}} = 0$) and sublattice potential ($\lambda_\mathrm{v} = 0$) . The intrinsic SOC is fixed at 0.06$t$.}\label{fig:bandstructure1}
		\end{center}
	\end{figure}
	\begin{figure}[h]
		\begin{center}
			\begin{subfigure}[b]{0.235\textwidth}
				\includegraphics[width=\textwidth]{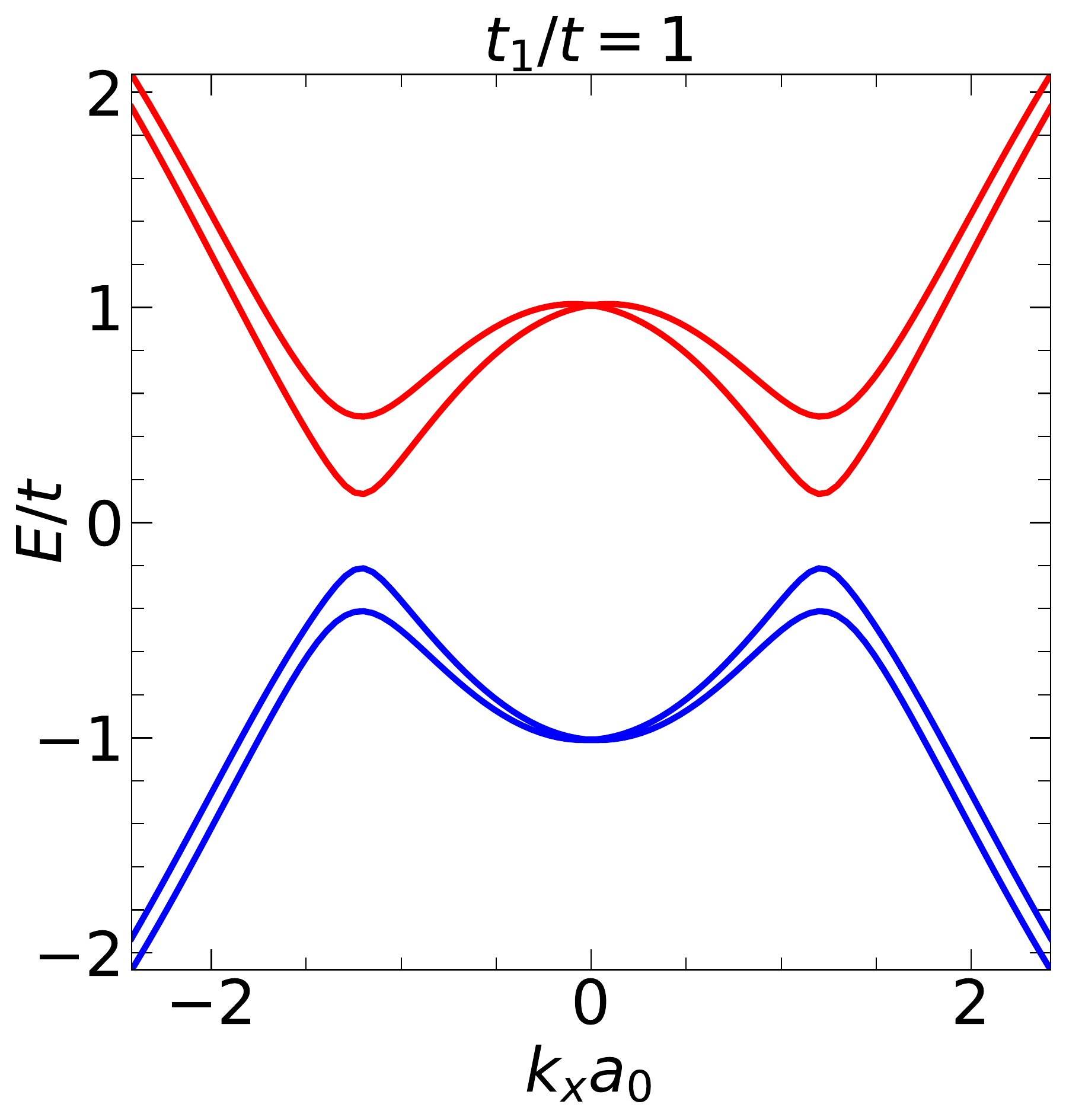}
				\subcaption{}\label{band1}
			\end{subfigure}
			\begin{subfigure}[b]{0.235\textwidth}
				\includegraphics[width=\textwidth]{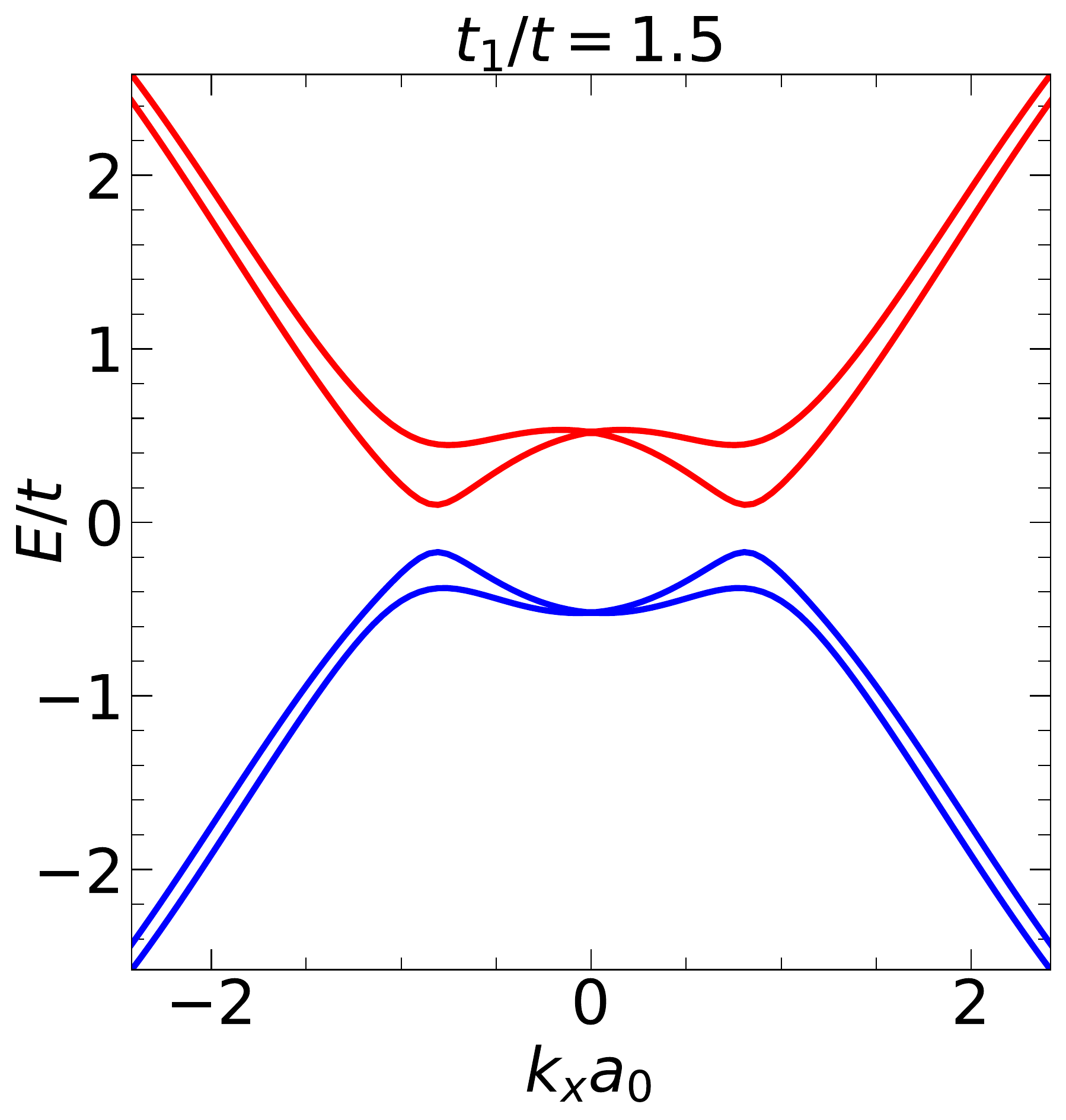}
				\subcaption{}\label{band2}
			\end{subfigure}
			\begin{subfigure}[b]{0.235\textwidth}
				\includegraphics[width=\textwidth]{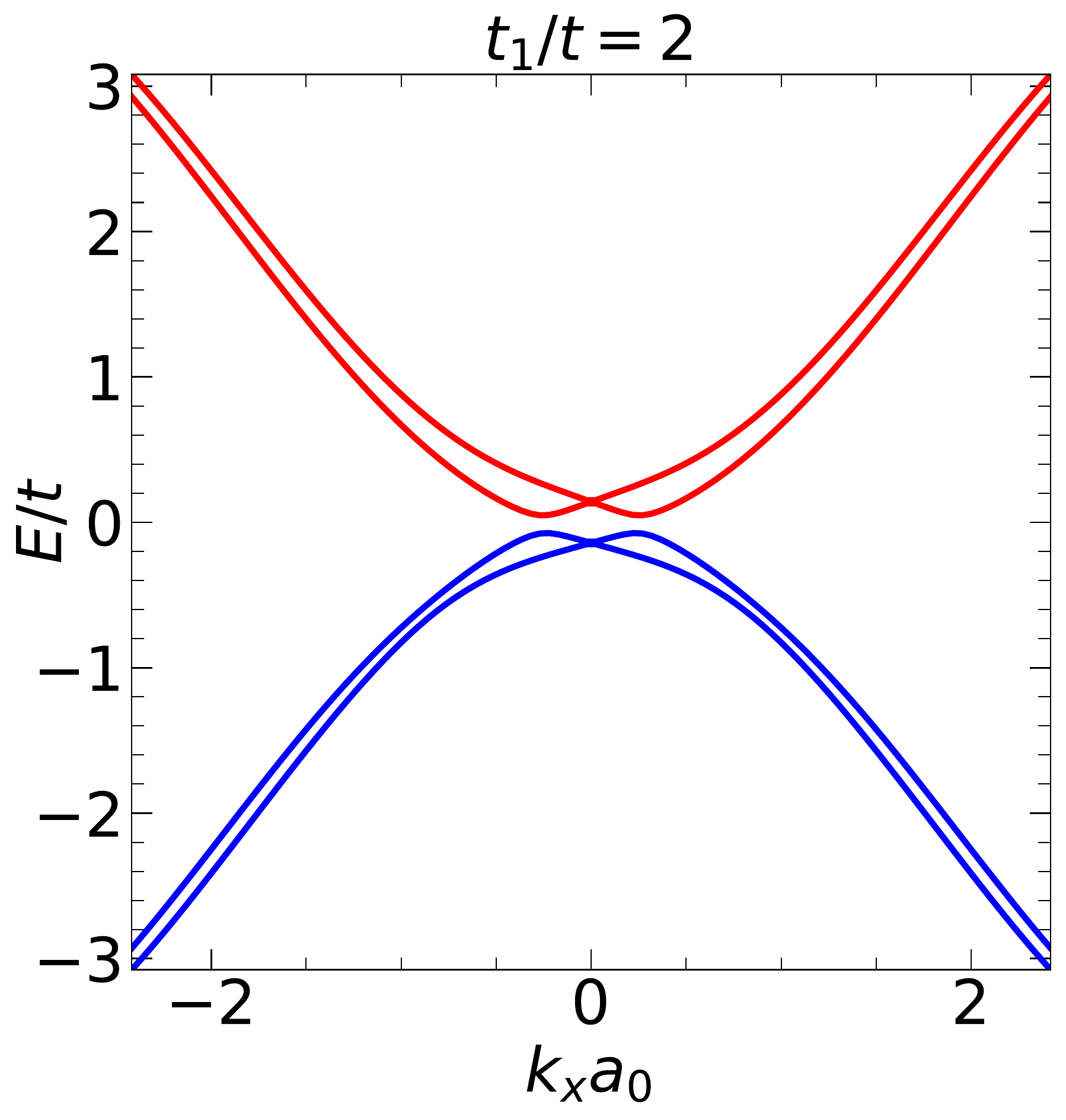}
				\subcaption{}\label{band3}
			\end{subfigure}
			\begin{subfigure}[b]{0.235\textwidth}
				\includegraphics[width=\textwidth]{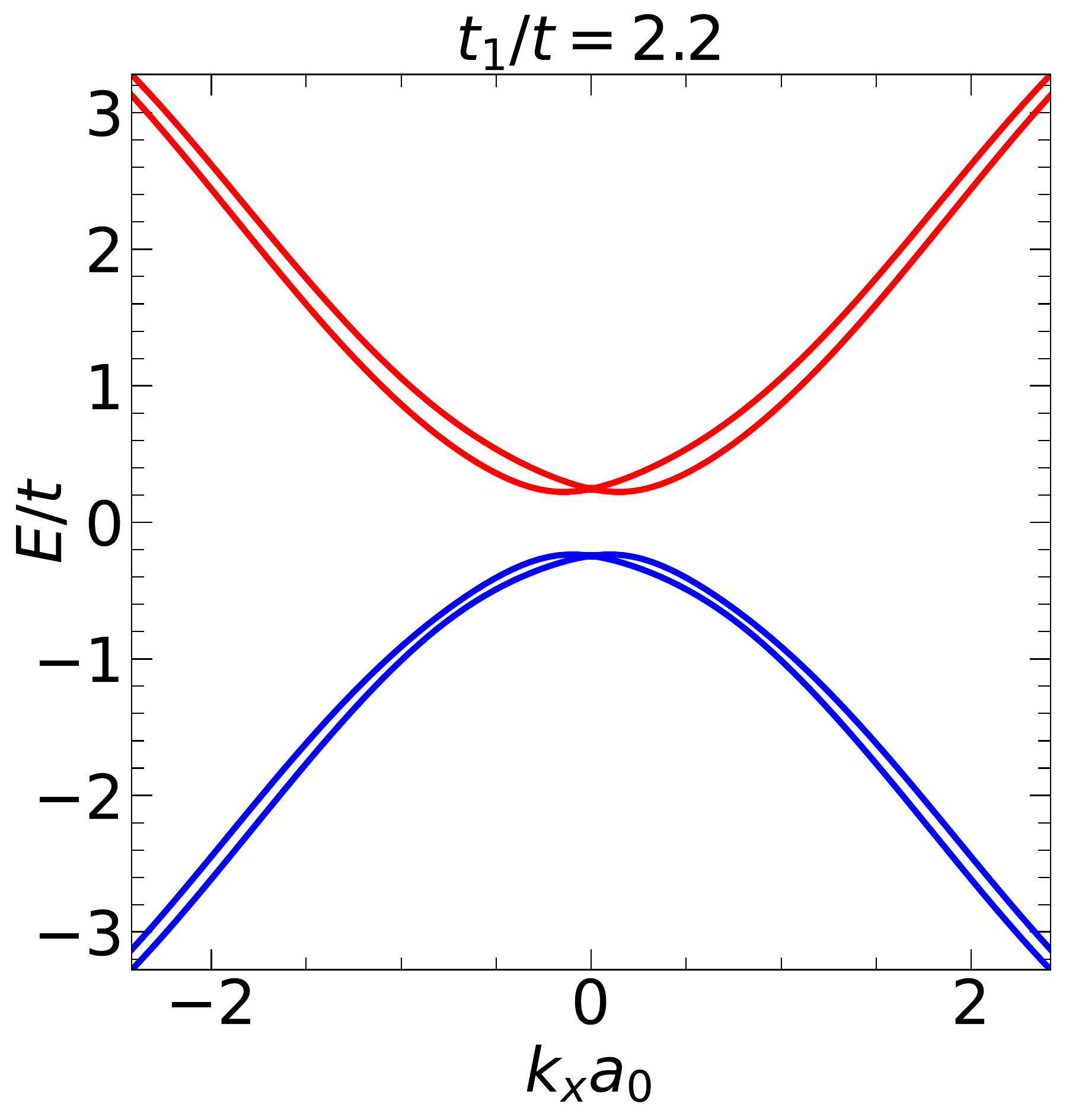}
				\subcaption{}\label{band4}
			\end{subfigure}
			\caption{\raggedright The energy band structure corresponding to the Hamiltonian in Eq. (\ref{ham_kspace}) along the $k_x$ axis ($k_y a_0$ is fixed at $2\pi/3$) are shown for (a) $t_1/t = 1$, (b) $t_1/t = 1.5$, (c) $t_1/t = 2$ and (d) $t_1/t = 2.2$. The other parameters are $\lambda_{\mathrm{R}} = 0.05t$, $\lambda_{\mathrm{SO}} = 0.06t$ and $\lambda_\mathrm{v} = 0.1t$. The $\lambda_{\mathrm{R}}$ (together with $\lambda_\mathrm{v}$) term makes the bands spin resolved and hence distinct. This is not the case in Fig. \ref{fig:bandstructure1}.}\label{fig:bandstructure}
		\end{center}
	\end{figure}
	The band structure of this Hamiltonian have been calculated numerically and presented in Fig. \ref{fig:bandstructure1} for various values of $t_1$ in absence of $\lambda_{\mathrm{R}}$ and $\lambda_\mathrm{v}$ ($\lambda_{\mathrm{R}}$ and $\lambda_{\mathrm{v}}$ are set to zero for now) so as to demonstrate the effects of the band deformation in presence of the spin dependent Haldane flux. The NNN hopping amplitude is fixed at 0.06$t$.
	For $t_1 = t$ (let us denote this as the Dirac case to distinguish it from the semi-Dirac case where the latter corresponds to $t_1 = 2t$), gaps open up at the two non-equivalent Dirac points, that is, at the $\mathbf{K}\;(2\pi/3\sqrt{3}a_0, 2\pi/3)$ and the $\mathbf{K}^\prime\;(-2\pi/3\sqrt{3}a_0, 2\pi/3)$ points as depicted in Fig. \ref{band5}. It should be noted that the band dispersion for the spin-$\uparrow$ and spin-$\downarrow$ electrons are identical. There are, in fact two conduction and two valence bands, one for each spin. However, in the absence of the Rashba and the Semenoff terms, these bands are degenerate. As $t_1$ is tuned to larger values, these two points come closer to each other and the band gap decreases. When $t_1$ becomes equal to $2t$, that is the semi-Dirac limit, the gap vanishes at the $\mathbf{M}\;(0,2\pi/3a_0)$ point (see Fig. \ref{band7}). Eventually when $t_1$ exceeds $2t$, the gap again opens up in the spectrum. The nature of the two gaps are different, and which is what we show below.
	
	Now if we include the Rashba coupling ($\lambda_{\mathrm{R}}$) and the sublattice potential ($\lambda_{\mathrm{v}}$), then the bands corresponding to the $\uparrow$-spins or $\downarrow$-spins become distinct as shown in Fig. \ref{fig:bandstructure}. Hence, in this case we see a pair of conduction and valence bands for each spin. It should be noted that below the semi-Dirac limit, that is, for $t_1 < 2t$, there exists a certain value of $\lambda_{\mathrm{R}}$ and $\lambda_\mathrm{v}$, where the gap vanishes completely and the bands touch (semi-metallic state). With further increase in the value of $\lambda_{\mathrm{R}}$ and $\lambda_\mathrm{v}$, the gap opens up again. However, such opening and closing of the spectral gaps are different in the semi-Dirac limit ($t_1=2t$), where there is no gap for $\lambda_{\mathrm{R}}$ and $\lambda_\mathrm{v}$ to be both zero. With the increase of $\lambda_{\mathrm{R}}$ and $\lambda_\mathrm{v}$, the gap opens up, but unlike the previous case, no closing of the band gap is observed in the dispersion spectrum. For $t_1>2t$, the dispersion is always gapped for any finite value of $\lambda_{\mathrm{R}}$ or $\lambda_\mathrm{v}$.

	\section{Edge states}\label{sec:edgestates}
	
\begin{figure}[h]
		\begin{center}
			\begin{subfigure}[b]{0.235\textwidth}
				\includegraphics[width=\textwidth]{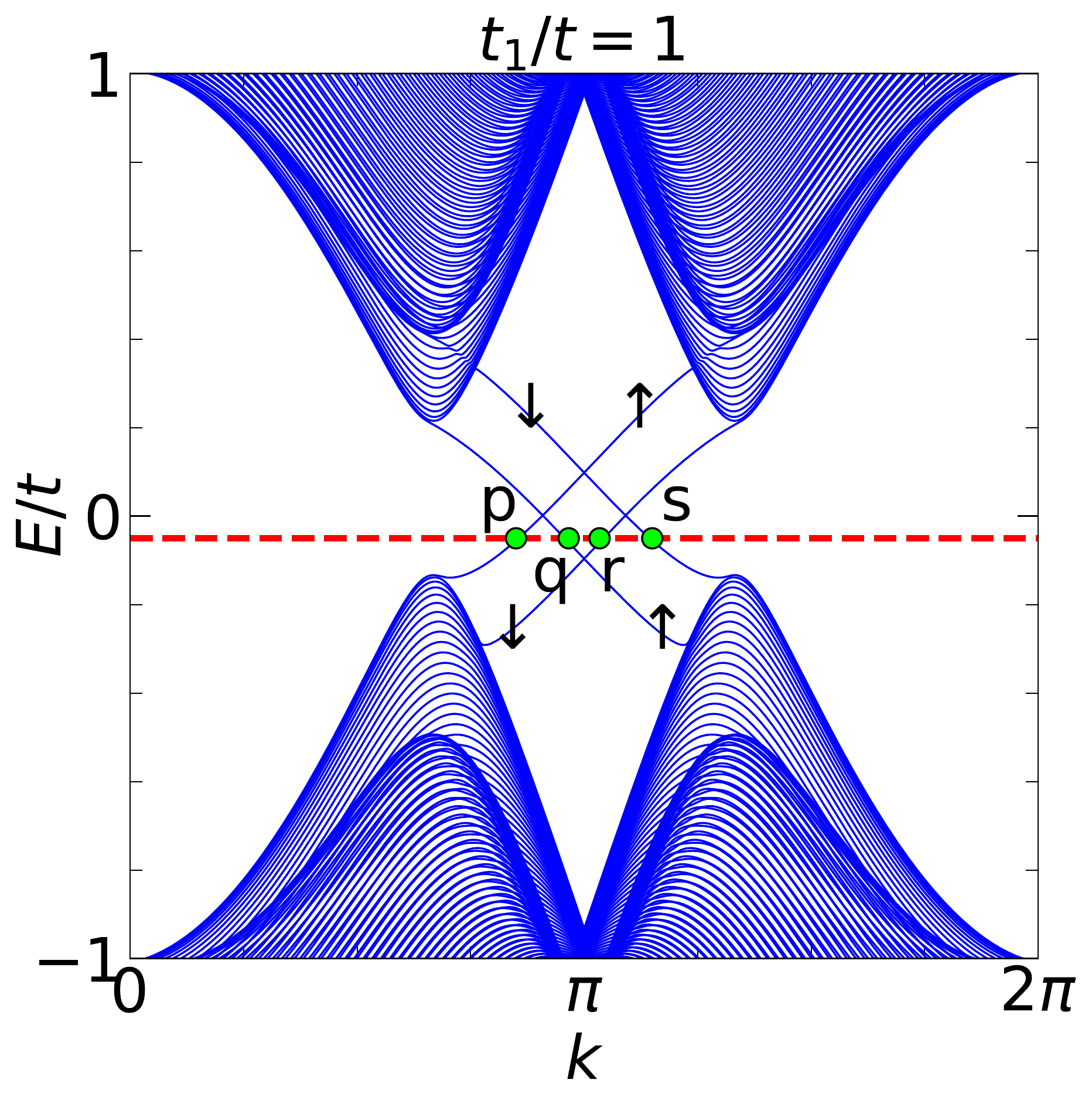}
				\subcaption{}\label{fig:edge1}
			\end{subfigure}
			\begin{subfigure}[b]{0.235\textwidth}
				\includegraphics[width=\textwidth]{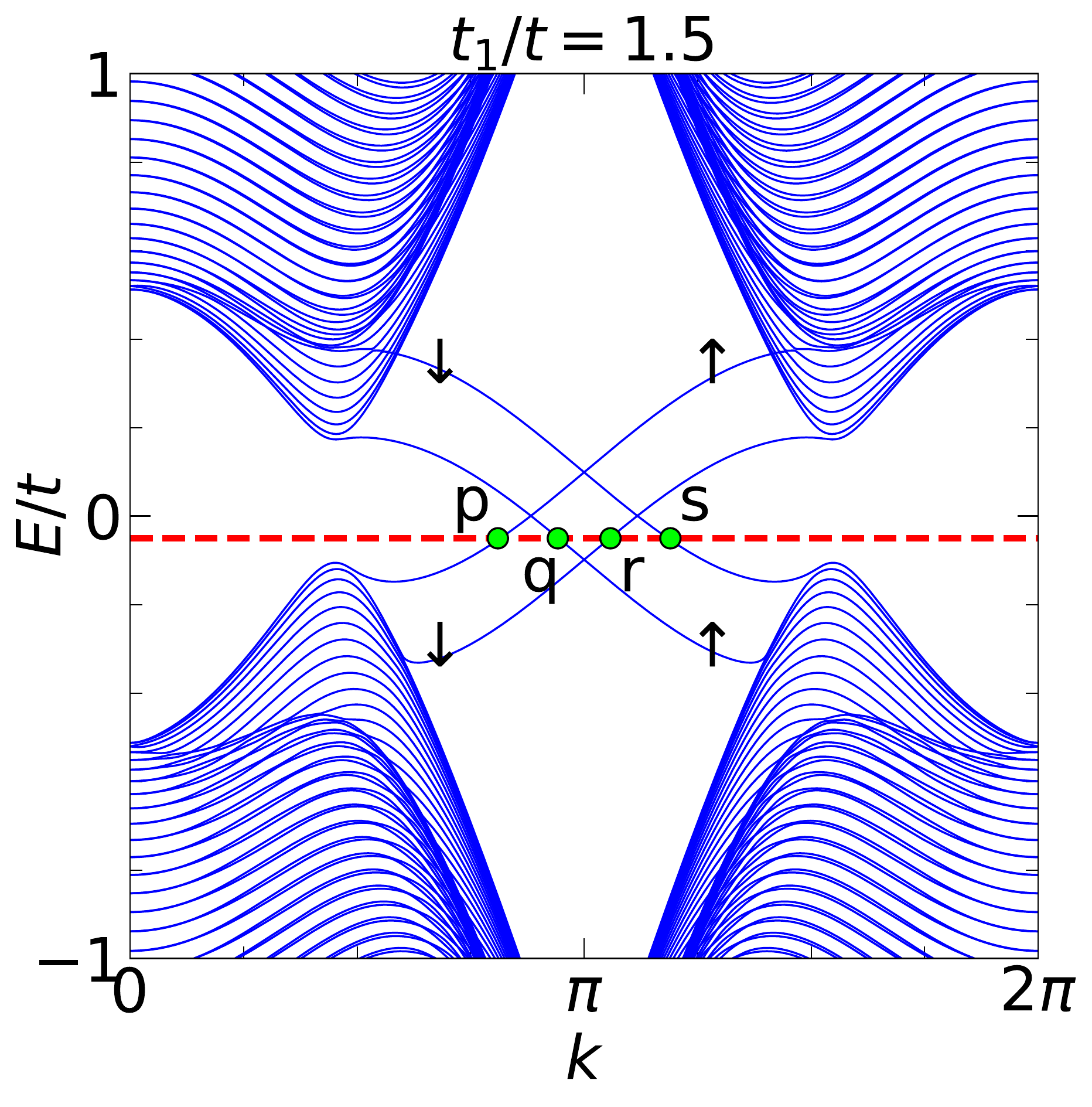}
				\subcaption{}\label{fig:edge2}
			\end{subfigure}
			\begin{subfigure}[b]{0.235\textwidth}
				\includegraphics[width=\textwidth]{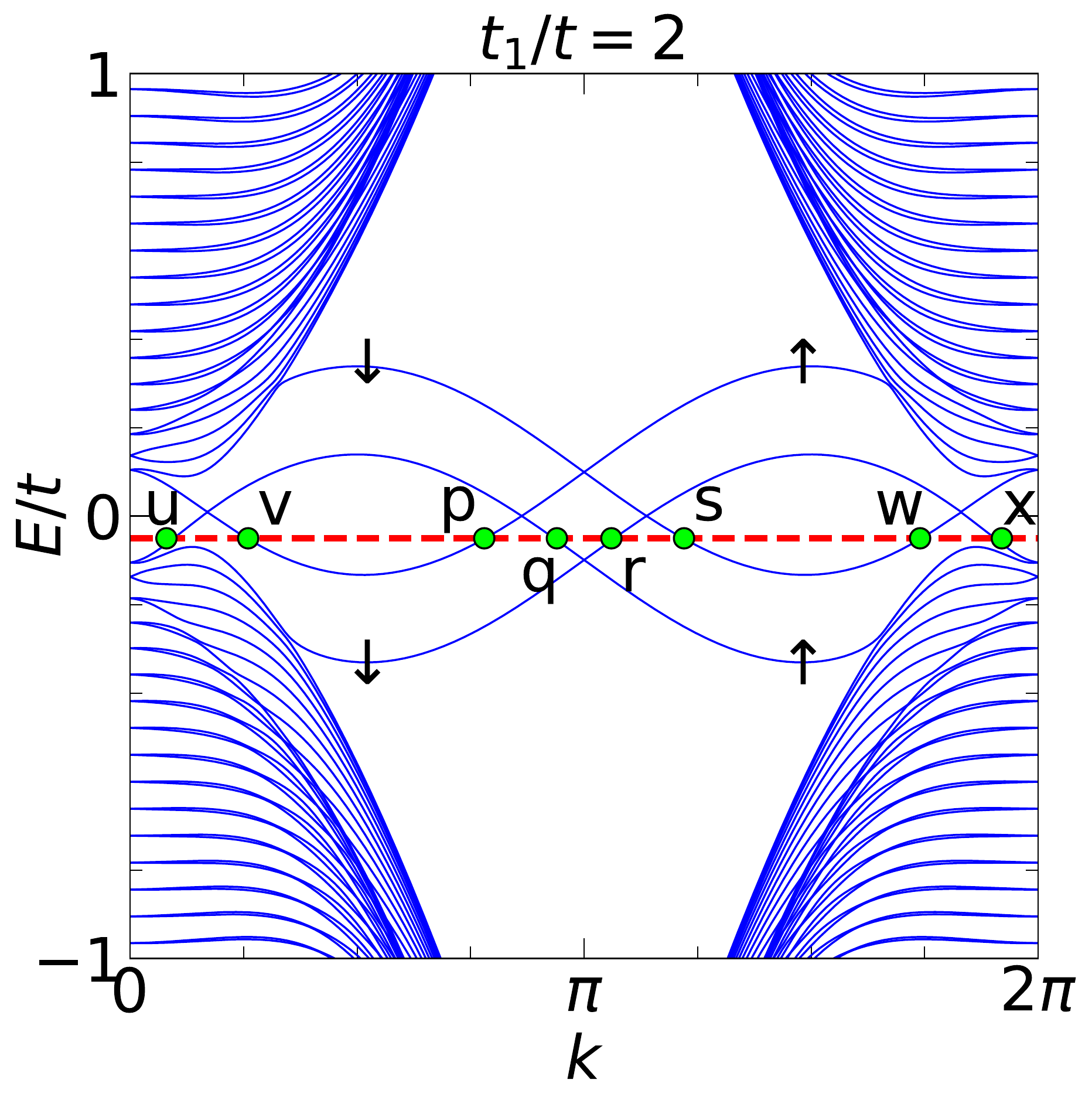}
				\subcaption{}\label{fig:edge3}
			\end{subfigure}
			\begin{subfigure}[b]{0.235\textwidth}
				\includegraphics[width=\textwidth]{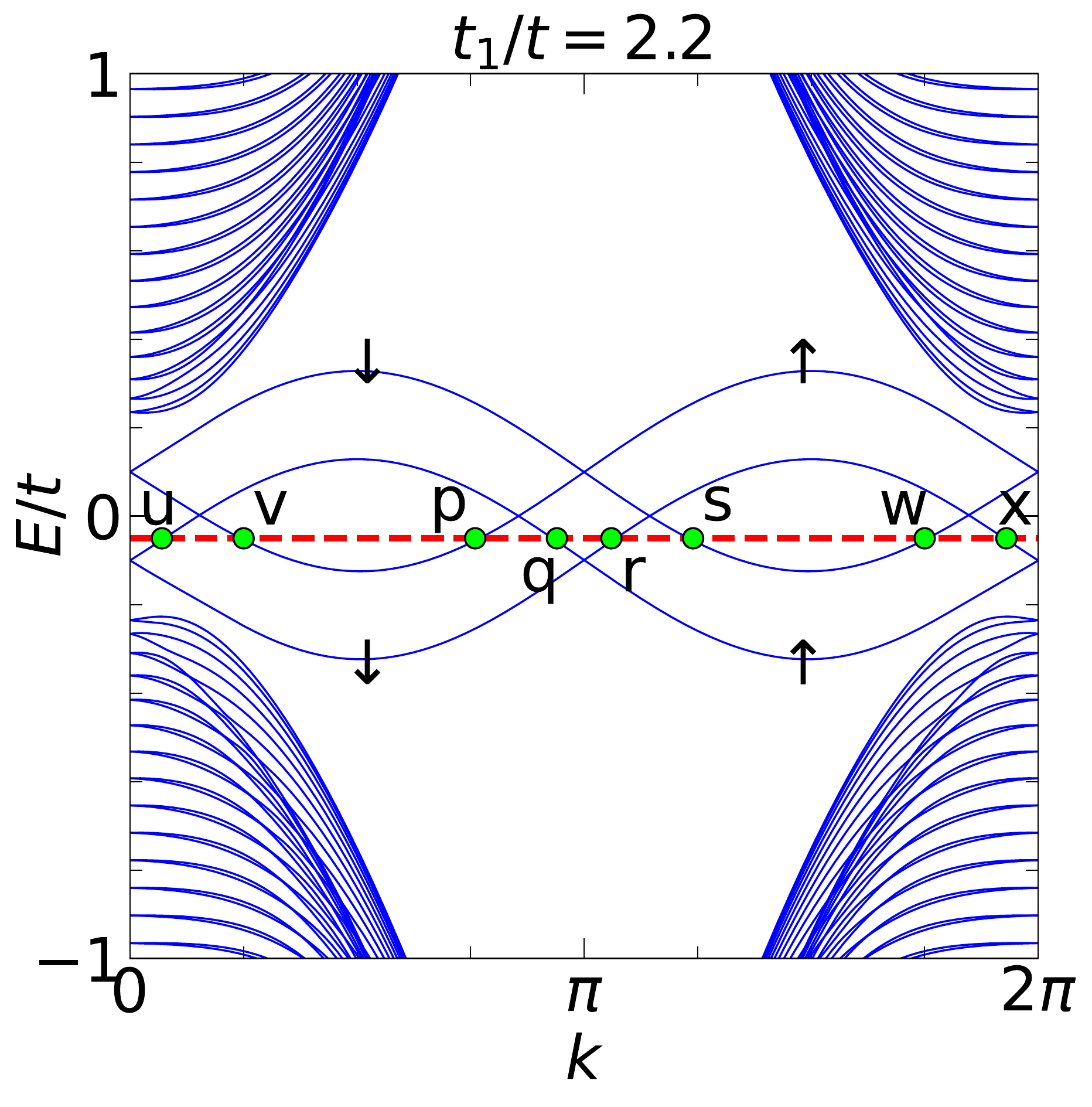}
				\subcaption{}\label{fig:edge4}
			\end{subfigure}
			\begin{subfigure}[b]{0.2\textwidth}
				\includegraphics[width=\textwidth]{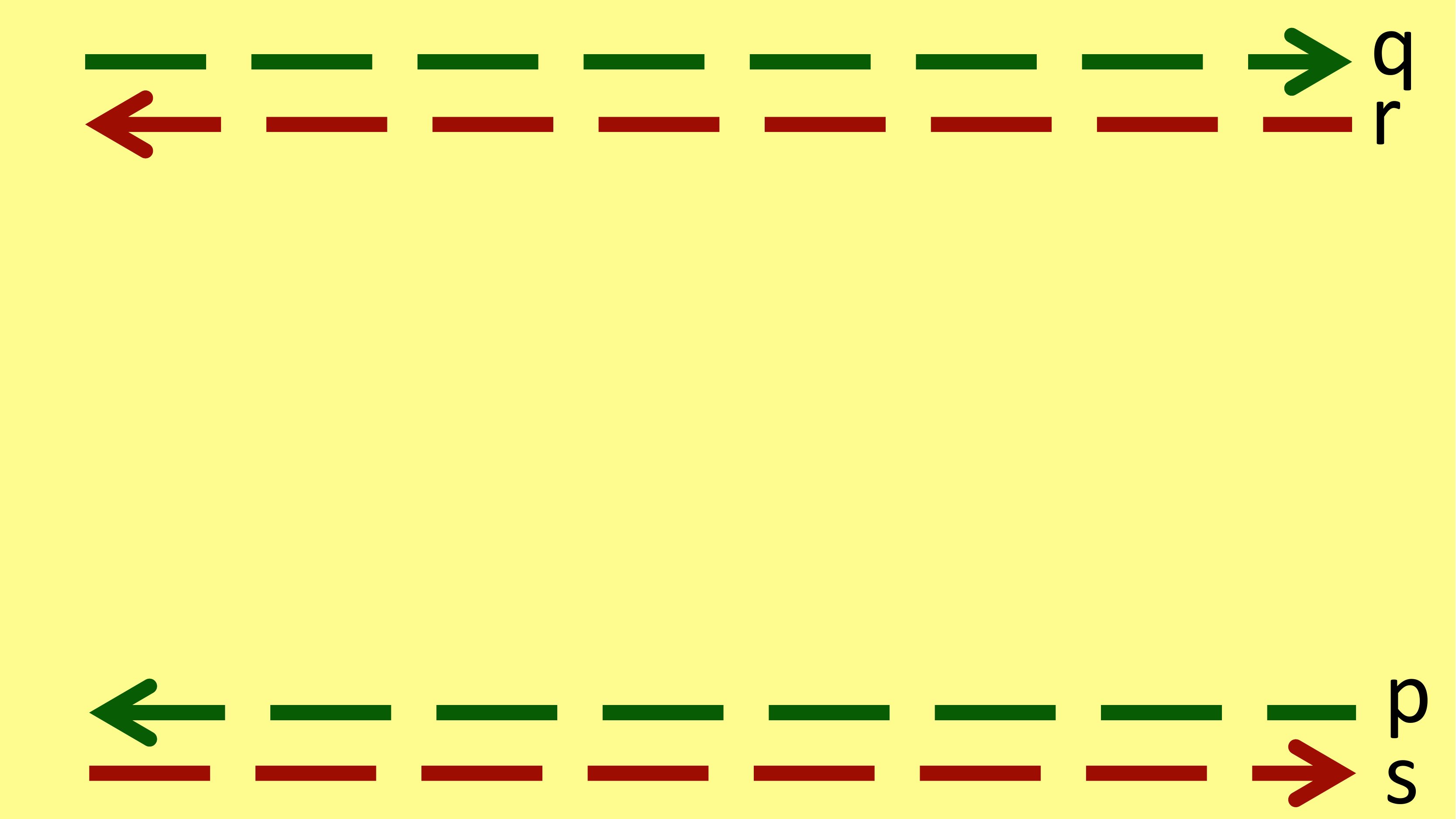}
				\subcaption{}\label{ribbon1}
			\end{subfigure}
			\begin{subfigure}[b]{0.2\textwidth}
				\includegraphics[width=\textwidth]{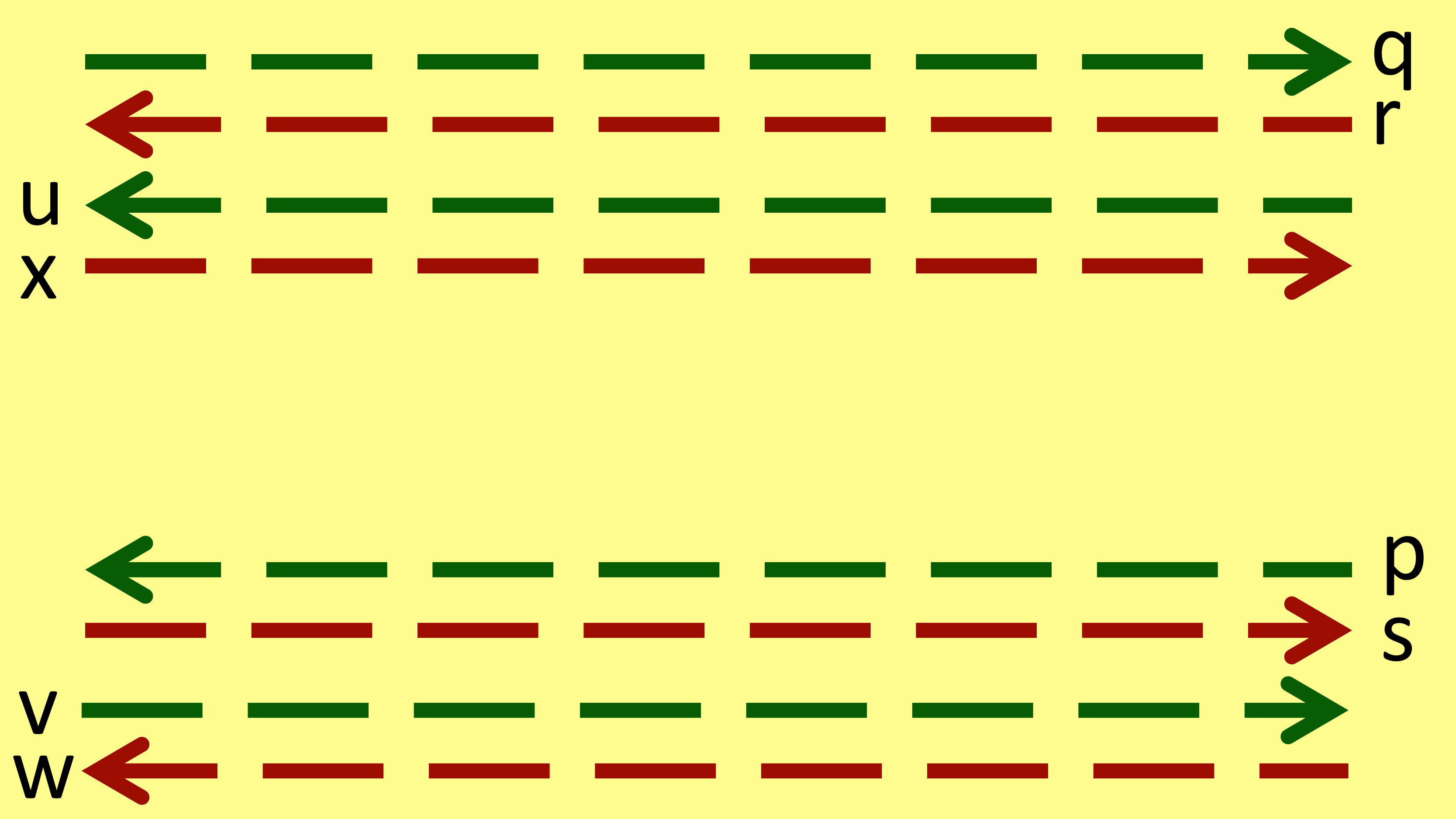}
				\subcaption{}\label{ribbon2}
			\end{subfigure}
		\end{center}
		\caption{\raggedright
				 The energy spectra  as a function of the dimensionless momentum $k$ (here $k$ denotes $\sqrt{3}a_0 k_{x}$) are shown for (a) $t_1/t = 1$, (b) $t_1/t = 1.5$, (c) $t_1/t = 2$ and (d) $t_1/t = 2.2$. In each figure, the edge modes corresponding to the spin-$\uparrow$ and spin-$\downarrow$ electrons are also depicted. The green dots in each figure signify the intersection of the edge states with the Fermi energy $E_F$ (shown via the red dashed line). A schematic diagram of a part of the ribbon with the edge currents corresponding to these intersecting points are shown in (e) and (f) in yellow panels, where the direction of the currents along the edges are depicted by arrows. The arrows in green colour represent the currents corresponding to spin-$\uparrow$ electrons, while those in red colour represent the same for spin-$\downarrow$ electrons. Through out the calculations $N$ ($D(N) = 149a_0$), $\lambda_{\mathrm{SO}}$, $\lambda_\mathrm{v}$ and $\lambda_{\mathrm{R}}$ are kept fixed at $100$, $0.06t$, $0.1t$ and $0.05t$ respectively.}\label{fig:edgestates}
	\end{figure}

	To understand the distinction between the spectral gaps observed for finite values of $\lambda_{\mathrm{R}}$ and $\lambda_\mathrm{v}$ ($\lambda_{\mathrm{SO}}$ is always taken to be non-zero in our discussion) corresponding to the Dirac ($t_1=t$) and the semi-Dirac ($t_1 = 2t$) cases, we ascertain the existence (or absence) of the edge states. To achieve this, we consider a finite strip of the system which breaks the periodicity along a particular direction, but allows translational invariance  along the perpendicular direction. Such a system is known as semi-infinite ribbon, which we have taken to be finite along the $y$-direction and infinite along the $x$-direction. We label the sites as A$_1$, B$_1$, A$_2$, B$_2$, .... A$_N$, B$_N$ etc. along the $y$-direction. Owing to translational invariance is the $x$-direction, we can Fourier transform the operators only in the $x$-direction, namely, use $c_{x, y}^\dagger = \sum_{k} e^{ikx}c_{k, y}^\dagger$. Using such a transformation on the Hamiltonian in Eq. (\ref{ham1}), we arrive at the following system of eigenvalue equations for the wave functions,
	
	\begin{equation}\label{edge1}
		\begin{aligned}
			E_{k} &a_{k, n} =\left[ t\left\{1+ e^{(-1)^{n+1} ik} \right\}b_{k, n} + t_1 b_{k, n+1} \right]s_0\\ & +2\lambda_{\mathrm{SO}}\left[ a_{k, n} \sin k  + e^{(-1)^{n+1}\frac{ik}{2}}\times \right.\\ &\left. \sin \frac{k}{2} \, \{a_{k, n-1} + a_{k, n+1} \} \right] s_z + \lambda_\mathrm{v} a_{k, n} s_0 + \\
			&i\lambda_\mathrm{R}\left[ \left\{ -\frac{1}{2}\left( 1 + e^{(-1)^{n+1} ik}\right)b_{k, n} + b_{k, n+1} \right\}s_y \right. \\ 
			& \left. -\left\{(-1)^n \frac{\sqrt{3}}{2}\left( 1 - e^{(-1)^{n+1}ik} \right)b_{k, n}\right\}s_x \right] 
		\end{aligned}
	\end{equation}
	\begin{equation}\label{edge2}
		\begin{aligned}
			E_{k} &b_{k, n} =\left[ t\left\{ 1+ e^{(-1)^{n} ik} \right\}a_{k, n} + t_1 a_{k, n-1} \right]s_0\\ & +2\lambda_{\mathrm{SO}}\left[ b_{k, n} \cos k + e^{(-1)^{n} \frac{ik}{2}}\times \right.\\ & \left. \cos \frac{k}{2} \{a_{k, n-1} + a_{k, n+1}\} \right] s_z - \lambda_\mathrm{v} b_{k, n}s_0 +\\
			&i\lambda_\mathrm{R}\left[ \left\{ \frac{1}{2}\left( 1 + e^{(-1)^{n} ik}\right)b_{k, n} + b_{k, n-1} \right\}s_y \right. \\ 
			& \left. -\left\{(-1)^{n+1} \frac{\sqrt{3}}{2}\left( 1 - e^{(-1)^{n}ik} \right)a_{k, n}\right\}s_x \right]
		\end{aligned}
	\end{equation}
	where $n$ stands for the $n$-th sublattice which is an integer in the range $[1:N]$ with $N$ being total number of unit cells along the $y$-direction and $k$ is the dimensionless momentum along the $x$-direction, such that, $k = \sqrt{3} a_0 k_x$. Also $s_i\,(i=x,y,z)$ denote the Pauli matrices and $s_0$ is the 2$\times$2 identity matrix. In the above equations $a_{k, n}$ and $b_{k, n}$ are the coefficients of the wavefunction corresponding to the sublattices A and B respectively, with momentum $k$. By solving these two equations, one can get the energy band structure for the ribbon for any arbitrary value of $t_1$. The width of the ribbon, $D$ can be written in terms of $N$, given by, $D(N) = a_0 \left(\frac{3N}{2} - 1\right)$. In our work, we have used $N = 100$ and hence the width of the ribbon is 149$a_0$ along the $y$-direction. As can be seen from the Fig. \ref{fig:edgestates}, a pair of modes from the lower (valence) band crosses over to the upper (conduction) band, and another pair crosses in the opposite direction. The amplitudes of the wave functions of these modes decay exponentially into the bulk, with their values being maximum at the edges of the ribbon \cite{nakada1996, castroneto, basu2018}. It should be noted that the velocities have opposite signs (since the velocity is proportional to $\partial E/\partial k$) along these modes, which implies that electrons propagate in opposite directions along each edge. These edge states are called as helical states as opposed to chiral modes corresponding to the quantum Hall case. 
	
	Let us scrutinize Fig. \ref{fig:edgestates} in more details. In Fig. \ref{fig:edge1} we have shown the band structure corresponding to the Dirac case ($t_1 = t$), where the intersection of the edge states with the Fermi energy, $E_F$ (shown via a red dashed line near $E/t = 0$) are represented by the green dots labelled as $p$, $q$, $r$ and $s$. The edge currents (proportional to the velocities) at those points are shown in the yellow panel of Fig. \ref{ribbon1}, where the yellow panel represents a part of the ribbon. As can be seen that there are counter propagating edge currents along either edge of the ribbon, and these edge currents are spin resolved, that is, they are different for spin-$\uparrow$ and spin-$\downarrow$. Since the system possesses TRS, the quantized (charge) Hall conductivity vanishes, however there will be a non-zero spin Hall conductivity, should the Fermi energy lies in the bulk gap. This state is known as the quantum spin Hall (QSH) phase and denotes a topological state of matter, that is, distinct from the familiar quantum Hall state. These edge currents will exist as long as the hopping strength $t_1$ remains lower than $2t$ (the semi-Dirac limit).
	
	In Fig. \ref{fig:edge2}, we have shown the band structure for $t_1 = 1.5t$ which is similar to the case $t_1 = t$, except that the bulk band gap being less. So, similar helical edge currents are observed for $t_1 = 1.5t$ (Fig. \ref{ribbon1}). Now if we increase $t_1$ to $2t$ or even beyond that (Fig. \ref{fig:edge3} and \ref{fig:edge4}), we see that there are 4 additional intersections of the edge states with the Fermi energy, $E_F$ which are denoted by the points $u$, $v$, $w$ and $x$. Hence there are a total of two pairs of the edge modes along either edge of the ribbon (Fig. \ref{ribbon2}), where for both $\uparrow$ and $\downarrow$ spins, the edge currents flow in opposite directions. Therefore in this case, the spin Hall conductivity vanishes completely and the system is devoid of any topological property \cite{kane2013}. This phenomenon persists for $t_1>2t$. Thus the edge states in a finite ribbon endorses a vanishing of the QSH phase in the semi-Dirac limit. Thus the `\textit{bulk-boundary correspondence}' observed for the QSH phase ceases to exist in the semi-Dirac limit, signalling the presence of a topological phase transition occurring through the  critical point, $t_1 = 2t$.
	
	\section{The Phase diagram}\label{sec:phasediagram}
	\begin{figure}
		\centering
		\includegraphics[width=0.4\textwidth]{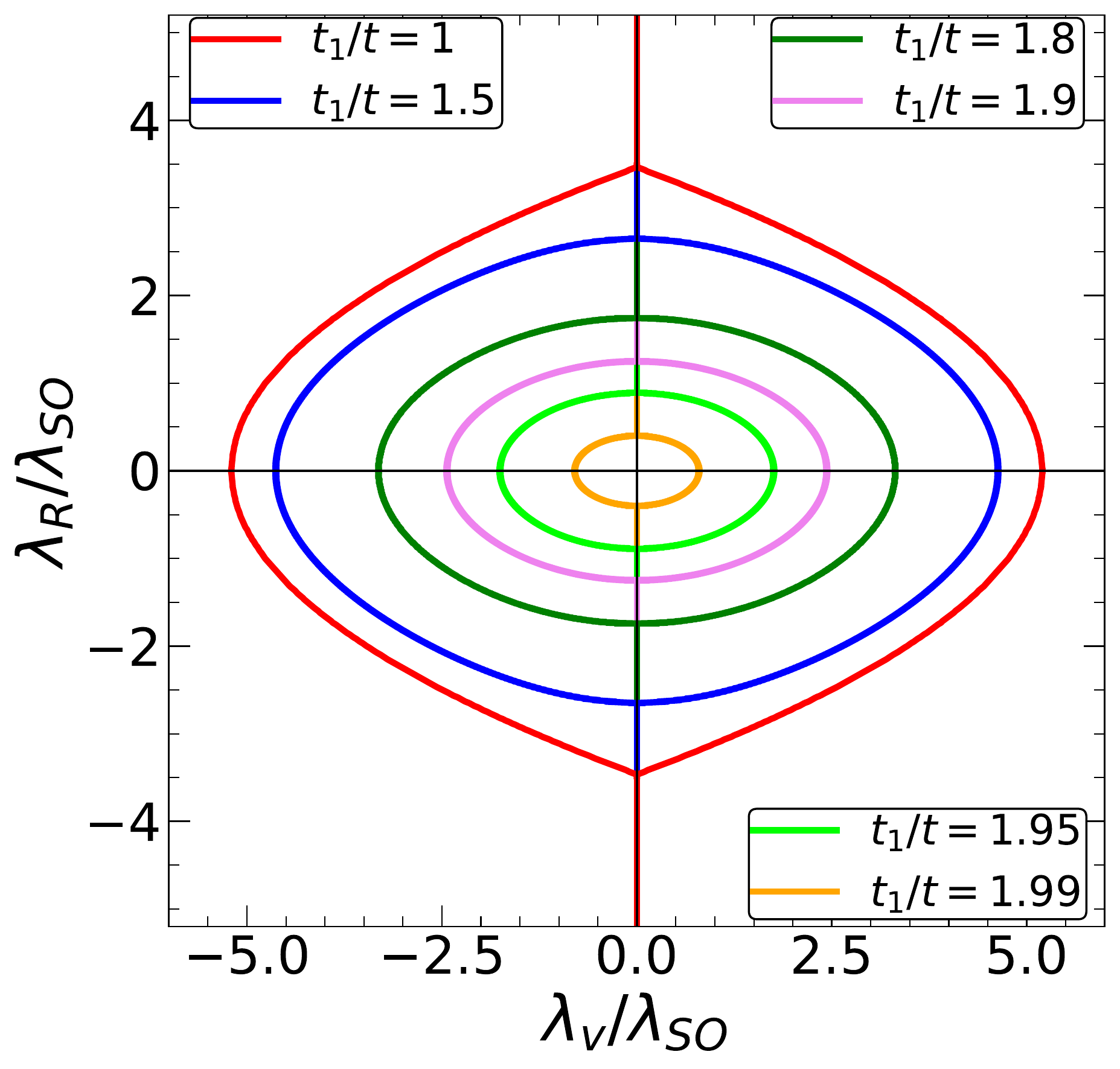}
		\caption{\raggedright The phase diagram of the system in the $\lambda_{\mathrm{R}}$-$\lambda_\mathrm{v}$ plane is shown for several values of $t_1$. The region enclosed by each curve represents the quantum spin Hall insulating phase where the $\mathbb{Z}_2$ index has a value 1, while the outside region denotes the trivial insulator where the $\mathbb{Z}_2$ index vanishes. The gradual vanishing of the area enclosed by each curve is shown.}\label{phase_diag}
	\end{figure}
	The quantum spin Hall (QSH) phase is characterized by a non-vanishing $\mathbb{Z}_2$ topological invariant (defined below). In this section, we show the variation of the $\mathbb{Z}_2$ invariant as we smoothly migrate from the Dirac to the semi-Dirac limit. For the calculation of this $\mathbb{Z}_2$ index, we have considered the Bloch wave functions, $u_i(\mathbf{k}_i)$ of the occupied bands corresponding to a pair of points $\mathbf{k}_1$ and $\mathbf{k}_2$ in the Brillouin zone. These two points denote the locations of the band extrema (minima for the conduction band and maximum for the valence band) in the BZ. Note that corresponding to the Dirac case, $\mathbf{k}_1$ and $\mathbf{k}_2$ denote the Dirac points $\mathbf{K}$ and $\mathbf{K}^\prime$ respectively, however they shift for $t_1\neq t$. The wave function at one of these points can be obtained by time reversing the wave function corresponding to the other one, that is, $\left| u_i(\mathbf{k}_1)\right> = \Theta\left| u_{i}(\mathbf{k}_2)\right>$ and vice versa where $\Theta$ denotes the time reversal operator. Since the Hamiltonian is time reversal invariant, so we can decompose the Hamiltonian, $H(\mathbf{k})$ and its corresponding occupied band wave functions, $|u_i(\mathbf{k})\rangle$ into even and odd subspaces. The even subspace  has the property that $\Theta\left| u_i(\mathbf{k})\right>$ is equivalent to $\left| u_i(\mathbf{k})\right>$ upto a $U(2)$ rotation. Whereas, the wave functions corresponding to the odd subspace has the property that the space spanned by $\Theta\left| u_i(\mathbf{k})\right>$ is orthogonal to that of $\left| u_i(\mathbf{k})\right>$. Now the $\mathbb{Z}_2$ invariant \cite{fu2006,kane2013} can be calculated by considering the momenta which belong to this odd subspace. We compute the expectation value of the time reversal operator between $|u_i(\mathbf{k})\rangle$ and $|u_j(\mathbf{k})\rangle$, namely, $\langle u_i(\mathbf{k} )| \Theta | u_j (\mathbf{k}) \rangle$. This yields a matrix which is antisymmetric. Hence we have,

	\begin{figure}
		\centering
		\includegraphics[width = 0.4\textwidth]{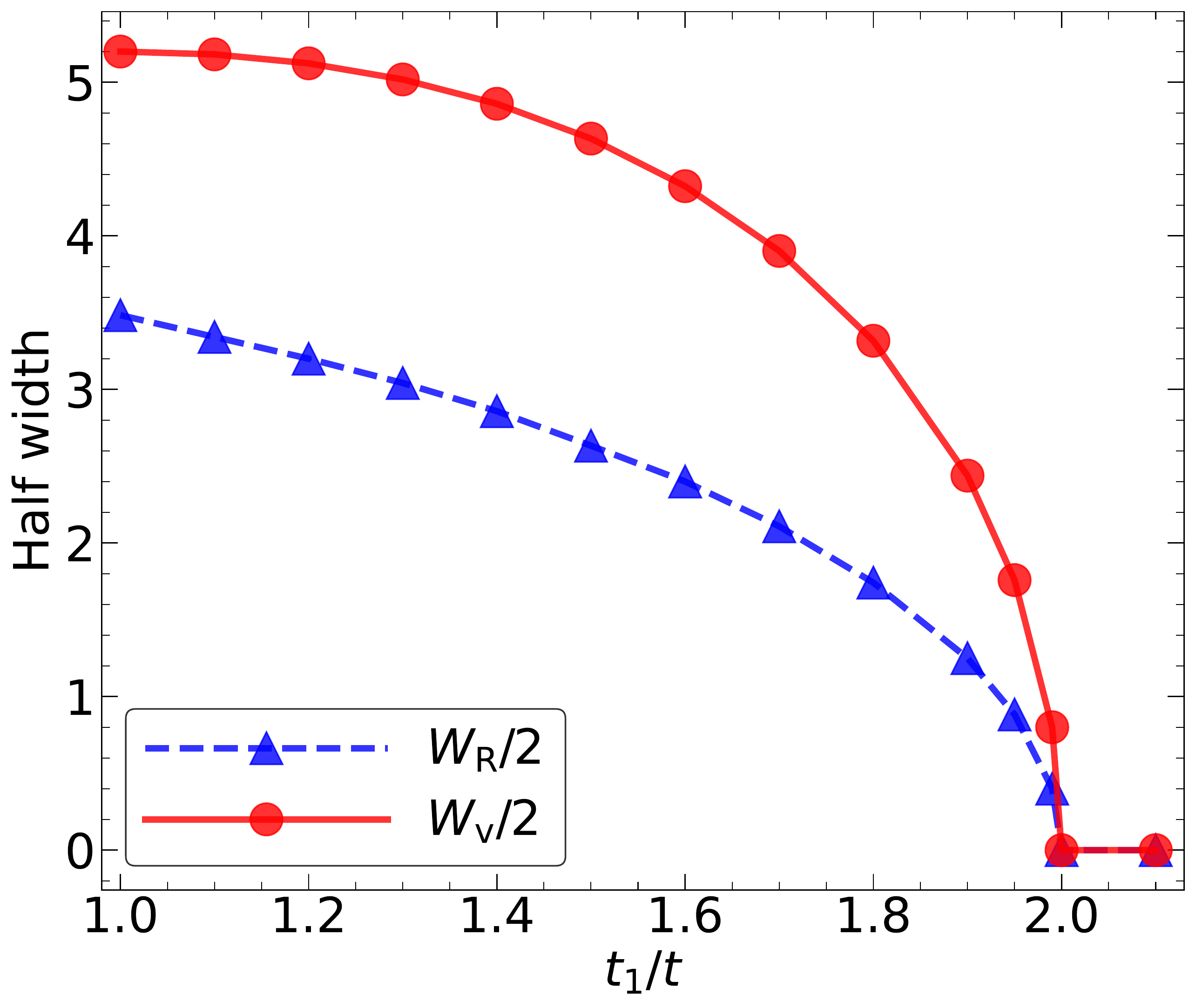}
		\caption{\raggedright The half width of the spin Hall insulating regime (refer to Fig. \ref{phase_diag}) along the $\lambda_\mathrm{v}/\lambda_{\mathrm{SO}}$ direction is shown by the red curve (denoted by $W_\mathrm{v}/2$), while along the $\lambda_{\mathrm{R}}/\lambda_{\mathrm{SO}}$ direction, it is shown by the blue curve (denoted by $W_\mathrm{R}/2$).}\label{width_in_pd}
	\end{figure}

	\begin{equation}
		\langle u_i(\mathbf{k} )| \Theta | u_j (\mathbf{k}) \rangle = \epsilon_{ij} P(\mathbf{k}),
	\end{equation}
	where $P(\mathbf{k})$ is the Pfaffian of the matrix \cite{aitken1939,krivo2016} defined as,
	\begin{equation}
		P(\mathbf{k}) = \mathrm{Pf}\left[ \langle u_i(\mathbf{k} )| \Theta | u_j (\mathbf{k}) \rangle \right]
	\end{equation} 
	For a $2\times2$ antisymmetric matrix $A_{ij}$, the Pfaffian picks up the off-diagonal component. Now the absolute value of this Pfaffian is unity in the even subspace, while it is zero in the odd subspace.  So we dissect the BZ into two halves such that the points $\mathbf{k}_1$ and $\mathbf{k}_2$ lie in different halves. Thus the $\mathbb{Z}_2$ index can be computed via performing the integral \cite{kanemele_2005},
	\begin{equation}
		\mathbb{Z}_2 = \frac{1}{2\pi i}\oint_C \mathrm{d}\mathbf{k} \cdot \boldsymbol{\nabla} \log\left( P(\mathbf{k}) + i\delta\right)
	\end{equation} 
	where $\delta$ is the convergence factor and the contour $C$ is the circumference of the halved BZ discussed above.

	In Fig. \ref{phase_diag} we have shown the phase diagram of the system as it is deformed from the Dirac to the semi-Dirac limit. For a given value of $t_1$, we have obtained the phase diagram in the $\lambda_\mathrm{R}$-$\lambda_\mathrm{v}$ plane. As can be seen that there are two regions in the phase diagram for all values of $t_1$ less than $2t$, comprising of the topological and the trivial phases. The region enclosed by each curve has the $\mathbb{Z}_2$ invariant equal to 1, while it vanishes outside the enclosed region. That means, the system shows non-trivial topology inside the enclosed region ($\mathbb{Z}_2 = 1$) and in the outside region, the system is trivial ($\mathbb{Z}_2 = 0$). It should be noted that the Dirac case ($t_1 = t$) yields the phase diagram obtained by Kane and Mele \cite{kanemele_2005} (red curve in Fig. \ref{phase_diag}) where the phase transition occurs at $\lambda_\mathrm{v} = 3\sqrt{3}\lambda_{\mathrm{SO}}$ for $\lambda_\mathrm{R} = 0$. Whereas, if we look at the phase diagrams for other values of $t_1$, that is intervening the Dirac and the semi-Dirac cases, the extent of the topological regime gradually shrinks. These results are consistent with the band structure plots (see Fig. \ref{fig:bandstructure1}). As the band gap decreases with increasing values of $t_1$, the phase transition from the topological to trivial phase now occurs for progressively lesser values of $\lambda_\mathrm{v}$ and $\lambda_{\mathrm{R}}$. In the semi-Dirac limit ($t_1 = 2t$), the system shows gapless band structure in the absence of both $\lambda_{\mathrm{R}}$ and $\lambda_\mathrm{v}$ and the $\mathbb{Z}_2$ number is identically equal to zero. Therefore, $t_1 = 2t$ denotes a critical point which demarcates the topological insulator from a band insulating phase. As the value of $t_1$ becomes larger than $2t$ (say $t_1 = 2.2t$), the system remains a band insulator.

	The shrinking of the half width (measured from the center in Fig. \ref{phase_diag}) along the $\lambda_{\mathrm{R}}$ axis ($\lambda_\mathrm{v} = 0$) of the non-trivial insulating region in the phase diagram is shown in Fig. \ref{width_in_pd} (the blue curve). As can be seen, the width ($W_\mathrm{R}/2$) is maximum at $t_1=t$ and has a value $2\sqrt{3}\lambda_{\mathrm{SO}}$ (non-zero $W_\mathrm{R}$ implies $\mathbb{Z}_2=1$, while $W_\mathrm{R} = 0$ implies $\mathbb{Z}_2=0$), which was obtained by Kane and Mele \cite{kanemele_2005}. Now with increase in the value of $t_1$, $W_\mathrm{R}/2$ decreases slowly. However there is a rapid decrease in the width as $t_1$ approaches $2t$. $W_\mathrm{R}/2$ eventually vanishes at the critical point (semi-Dirac), namely, $t_1 = 2t$.
	In the same figure (Fig. \ref{width_in_pd}) we have also depicted the variation of the half width  along the $\lambda_\mathrm{v}$ direction ($\lambda_{\mathrm{R}} = 0$), namely $W_\mathrm{v}/2$ as a function of $t_1/t$. In this case also, the width $W_\mathrm{v}/2$ falls off slowly near $t_1 = t$ but as $t_1$ increases, $W_\mathrm{v}/2$ drops rapidly in the vicinity of $t_1 = 2t$. $W_\mathrm{v}/2$ vanishes at the critical point, $t_1 = 2t$ and continues to remain at zero for $t_1>2t$. The analytic form of the width, say for $W_\mathrm{v}/2$ as a function of $t_1/t$ can be obtained as,
	
	\begin{equation}
		\frac{W_\mathrm{v}}{2} = \left( 2+\frac{t_1}{t} \right)^{3/2} \sqrt{2-\frac{t_1}{t}}
	\end{equation}
	where the equation is valid for $t\leq t_1 < 2t$. From this equation one can get the half width in the phase diagram along the horizontal direction for $t_1 = t$ is $3\sqrt{3} \lambda_{\mathrm{SO}}$, which is a well known result in the Kane-Mele model \cite{kanemele_2005} and mentioned earlier. For $t_1 = 2t$, this width vanishes and the topological regime ceases to exist. 
	
	\section{Spin Hall conductivity}\label{sec:spinhallcond}
	In this section we present numerical calculations of the spin Hall conductivity for our system. In order to calculate the spin Hall conductivity, we first obtain the low energy form of the Hamiltonian in Eq. \ref{ham1} for different values of $t_1$. Such a low energy expansion of the Hamiltonian will be helpful for our purpose, which can be written in a compact notation as, 
	\begin{align}\label{low_ham}
		H(q_x, q_y) =& \zeta_x(q_x, q_y, t_1)\sigma_x s_0 + \zeta_y(q_x, q_y, t_1)\sigma_y s_0 + \nonumber\\
		&\gamma(q_x, q_y, t_1)\sigma_z s_z + \left[ \right. \nonumber\\
		& \left.\rho_x(q_x, q_y,t_1) \sigma_x + \rho_y(q_x, q_y,t_1) \sigma_y \right] (s_x - s_y)
	\end{align}
	where the coefficients $\zeta_x$, $\zeta_y$, $\gamma$ and  $\rho$ are functions of the momentum $\mathbf{q}$ (measured relative to the band extrema) and the hopping strength $t_1$. $s_x$, $s_y$ and $s_z$ are the 2$\times$2 Pauli matrices which represent the real spin of the electrons, while the other 2$\times$2 Pauli matrices $\sigma_x$, $\sigma_y$ and $\sigma_z$ represent the sublattice degree of freedom and $s_0$ is the 2$\times$2 identity matrix. The full tight binding forms for $\zeta_x$, $\zeta_y$, $\gamma$ and $\rho$ are presented in section \ref{sec:modelhamiltonian}. It should be noted that the low energy expansions are done at different points in the Brillouin zone for different values of $t_1$, since the band minima shift as we change $t_1$ relative to $t$ (see Fig. \ref{fig:bandstructure1}). For example, at $t_1 = t$, the band minima occur at $\mathbf{K}\;(2\pi/3\sqrt{3}a_0, 2\pi/3a_0)$ and at $\mathbf{K}^\prime\;(-2\pi/3\sqrt{3}a_0, 2\pi/3a_0)$ points, whereas at $t_1 = 2t$ the band minima occur at the $\mathbf{M}(0, 2\pi/3a_0)$ point in the BZ. Thus the approximations are done at locations in the BZ where the spectral gap is minimum for all values of $t_1$.
	
	\begin{figure}[h]
		\centering
		\includegraphics[width = 0.4\textwidth]{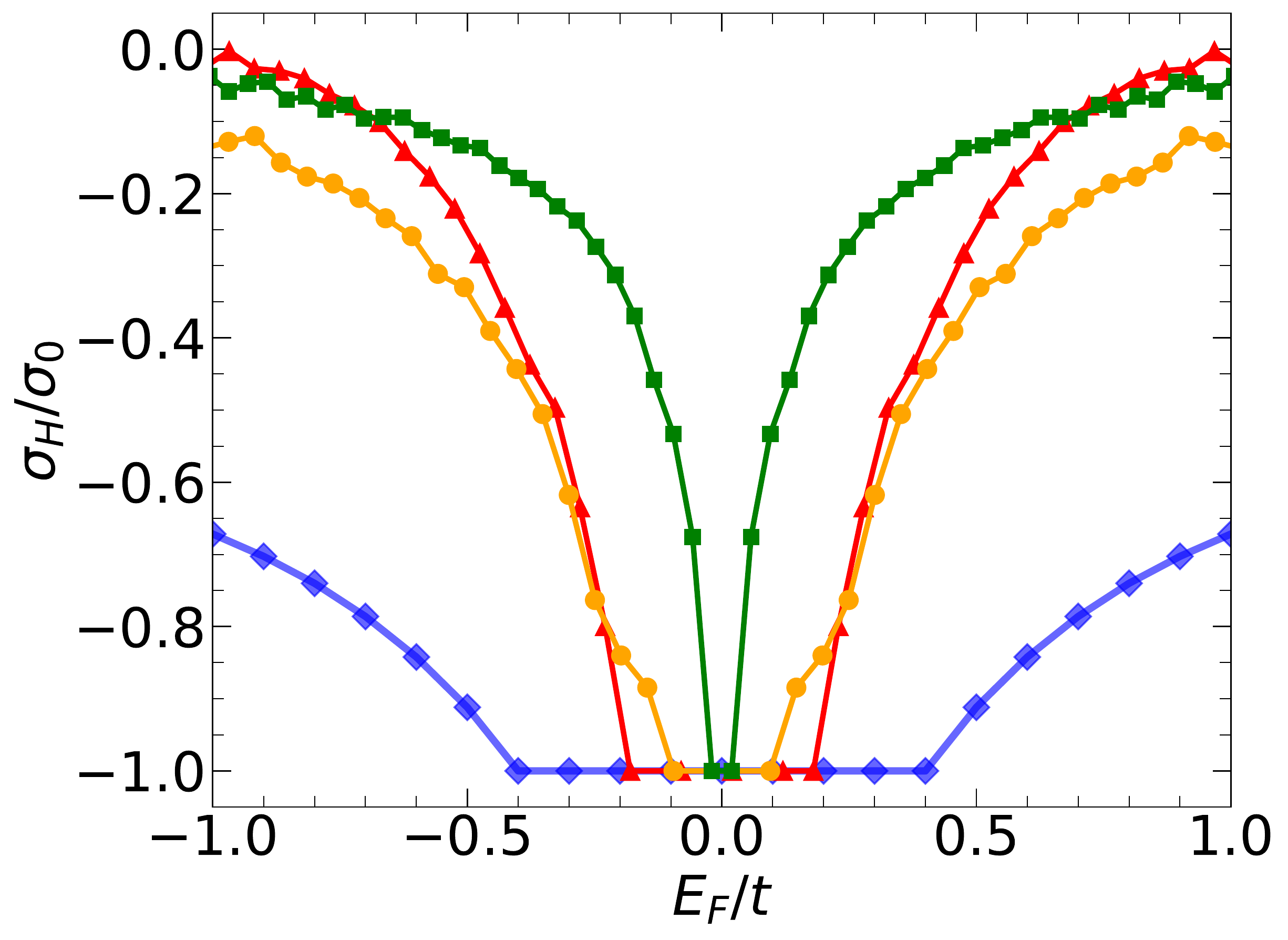}
		\caption{\raggedright The evolution of the spin Hall conductivity of the system for various values of $t_1$ as a function of the Fermi energy $E_F$. The plateau width gets gradually narrower, before vanishing at the semi-Dirac limit.}\label{fig:hallcond}
	\end{figure}
	In order to calculate the Hall conductivity we have used the following formula \cite{taguchi2020,rimondi2004,tanaka2008},
	
	\begin{equation}\label{hall_cond}
		\sigma_{H}^z = e\hbar \sum_\lambda \int \frac{d\mathbf{q}}{(2\pi)^2} f(E_n) \Omega^z_{xy}(\mathbf{q})
	\end{equation}
	where $f(E_n) = [e^{\beta(E_n - E_F)} + 1]^{-1}$ is the Fermi-Dirac distribution function corresponding to a given energy $E_n$ and $\beta$ is the inverse temperature. $\Omega_{xy}^z$ is the Berry curvature which can be evaluated from,
	\begin{equation}\label{berrycurv}
		\Omega_{H}^z (\mathbf{q}) = i\sum_{m\neq n}^{ }\frac{\langle n|v_{sx}^z|m\rangle \langle m |v_y |n\rangle - (x\leftrightarrow y)}{(E_n(\mathbf{q}) - E_m(\mathbf{q}))^2}
	\end{equation}
	where $x\leftrightarrow y$ denotes interchanging $x$ and $y$ variables in the first term. Further, $\Omega_{H}^z (\mathbf{q})$ is the spin dependent Berry curvature, $\{| n \rangle, E_n\}$ denotes the eigensolutions corresponding to the Hamiltonian in Eq. \ref{low_ham}, $v_i = \frac{1}{\hbar}\partial H/\partial k_i$ is the velocity operator and $v_{si}^z =\frac{1}{2} \{ v_i, \frac{\hbar}{2} s_z \sigma_0 \}$ is the velocity operator corresponding to the spin current. The integral is performed over the states which are occupied at a given value of the Fermi energy. Using Eqs. \ref{low_ham}, \ref{hall_cond} and \ref{berrycurv}, one can compute the spin Hall conductivity numerically as a function of $E_F$ for various values of $t_1$ as shown in Fig. \ref{fig:hallcond}. We can see that as long as the Fermi energy lies in the bulk gap, the Hall conductivity possesses a plateau and hence quantized in unit of $\sigma_0$ ($=e/2\pi$) \cite{kanemele_2005_2}. The spin Hall conductivity however decreases rapidly as the Fermi energy moves away from the gapped region. With increase in the value of $t_1$, the quantized plateau stays at $e/2\pi$, except that the plateau width decreases, which eventually vanishes at $t_1 = 2t$. Thus a vanishing of the QSH phase indeed occurs because of the band deformation induced by the hopping anisotropy as discussed throughout the paper. The above scenario also implies that $\mathbb{Z}_2$ invariant vanishes in the semi-Dirac limit.
	
	\section{Conclusion}\label{sec:conclusion}
	
	We have investigated the evolution of the electronic dispersion and the topological properties in the semi-Dirac Kane Mele model with the inclusion of a differential hopping between the neighbours on a honeycomb lattice. This causes deformation of the (spin resolved) bands which are evident from our band structure calculations. The electronic band dispersion shows the band minima originally located at the Dirac points to migrate towards each other and finally merge at the $\mathbf{M}$ point, where the gap vanishes completely in the semi-Dirac limit. The model shows a topological phase transition from a non-trivial spin Hall insulating phase to a trivial band insulator at the semi-Dirac limit. Further, such a topological phase transition is adequately supported via computing the helical edge modes, $\mathbb{Z}_2$ invariant, phase diagram and the spin Hall conductivity.

\end{document}